\documentclass[conference,compsoc]{IEEEtran}

\ifCLASSOPTIONcompsoc
  \usepackage[nocompress]{cite}
\else

  \usepackage{cite}
\fi

\ifCLASSINFOpdf
  \usepackage[pdftex]{graphicx}
\else
  \usepackage[dvips]{graphicx}
\fi

\usepackage{amsmath}
\ifCLASSOPTIONcompsoc
 \usepackage[caption=false,font=footnotesize,labelfont=sf,textfont=sf]{subfig}
\else
 \usepackage[caption=false,font=footnotesize]{subfig}
\fi

\usepackage{amsmath}
\usepackage{amsthm}
\usepackage{amssymb}
\usepackage[bookmarks]{hyperref}
\hypersetup{hidelinks}
\usepackage{mathpartir}
\usepackage{stmaryrd}
\usepackage{tabularx}
\usepackage{color}

\theoremstyle{plain}
\newtheorem{thm}{Theorem}[section]

\newtheorem{lem}[thm]{Lemma}
\newtheorem{cor}[thm]{Corollary}
\theoremstyle{definition}
\newtheorem{dfn}[thm]{Definition}

\newenvironment{example*}{\begin{exa*}\normalfont}{\qed\end{exa*}}

\newcommand{\NI}{\textit{non-interference}}
\newcommand{\cg}{CG}
\newcommand{\fg}{FG}

\newcommand{\fc}{FC}
\newcommand{\rfg}{$\mbox{\fg}^-$}

\newcommand{\update}[1]{#1}
\newcommand{\kw}[1]{\mathsf{#1}}
\newcommand{\efst}{\kw{fst}}
\newcommand{\esnd}{\kw{snd}}
\newcommand{\ecase}{\kw{case}}
\newcommand{\einl}{\kw{inl}}
\newcommand{\einr}{\kw{inr}}

\newcommand{\eret}{\kw{ret}}
\newcommand{\ebind}{\kw{bind}}
\newcommand{\elabel}[1]{\kw{Lb}_{#1}}
\newcommand{\etolabeled}{\kw{toLabeled}}
\newcommand{\eunlabel}{\kw{unlabel}}

\newcommand{\enew}{\kw{new}~}

\newcommand{\subtype}{\mathrel{<:}}

\newcommand{\utype}{\kw{A}}
\newcommand{\tbase}{\kw{b}}

\newcommand{\tlabeled}[2]{\kw{Labeled}\;#1\;#2}
\newcommand{\tslio}[3]{\mathbb{C}\;#1\;#2\;#3}
\newcommand{\tunit}{\kw{unit}}
\newcommand{\tbool}{\kw{bool}}

\newcommand{\tref}{\kw{ref}}

\newcommand{\rname}[1]{\;\mbox{\small #1}}

\newcommand{\llabel}{\ell}

\newcommand{\ljoin}{\sqcup}

\newcommand{\lbelow}{\sqsubseteq}

\newcommand{\fto}{\overset{\llabel_e}\to}

\newcommand{\ftoP}{\overset{\llabel_e'}\to}

\newcommand{\pc}{\mathit{pc}}
\newcommand{\dom}{\mathtt{dom}}

\newcommand{\lattice}{\mathcal{L}}

\newcommand{\union}{\cup}

\newcommand{\transFR}[1]{\ensuremath{\llbracket #1 \rrbracket}}
\newcommand{\trans}[1]{\transFR {#1}}

\newcommand{\transFC}[1]{\ensuremath{\llparenthesis #1 \rrparenthesis}}

\newcommand{\coerce}{\ensuremath{\mathtt{coerce\_taint}}}

\newcommand{\Blrv}[1]{\ensuremath{\lceil #1 \rceil}_{V}^{\attacker}}
\newcommand{\Blre}[1]{\ensuremath{\lceil #1 \rceil}_{E}^{\attacker}}

\newcommand{\Ulrv}[1]{\ensuremath{\lfloor #1 \rfloor}_{V}}
\newcommand{\Ulre}[1]{\ensuremath{\lfloor #1 \rfloor}_{E}^{\pc}}
\newcommand{\lreu}[2]{\ensuremath{\lfloor #1 \rfloor}_{E}^{#2}}

\newcommand{\fcUlrvn}[1]{\ensuremath{\lfloor #1 \rfloor}_{V}^{\pb}}
\newcommand{\fclrvun}[2]{\ensuremath{\lfloor #1 \rfloor}_{V}^{#2}}
\newcommand{\fcUlren}[1]{\ensuremath{\lfloor #1 \rfloor}_{E}^{\pb}}
\newcommand{\fclreun}[2]{\ensuremath{\lfloor #1 \rfloor}_{E}^{#2}}

\newcommand{\world}{\mathit{W}}

\newcommand{\uWorld}{\theta}
\newcommand{\uWorlds}{{^s}\theta}

\newcommand{\loc}{\mathit{a}}

\newcommand{\val}{\mathit{v}}
\newcommand{\sval}{{^s}\mathit{v}}
\newcommand{\tval}{{^t}\mathit{v}}
\newcommand{\attacker}{\mathcal{A}}
\newcommand{\wExtends}{\sqsubseteq}
\newcommand{\wExtendsR}{\sqsupseteq}
\newcommand{\pb}{\hat{\beta}}
\newcommand{\heap}{\mathit{H}}

\newcommand{\wDefined}[1]{\overset{#1}\triangleright}

\newcommand{\ReducesTo}[1]{\Downarrow_{#1}}
\newcommand{\ReducesToF}[1]{\Downarrow_{#1}^{f}}
\newcommand{\proj}[1]{{\downarrow_{#1}}}
\newcommand{\vsubstp}{\gamma}
\newcommand{\vsubsts}{\delta}
\newcommand{\vsubstsS}{\delta^s}
\newcommand{\vsubstsT}{\delta^t}

\begin{document}

%\title{Type Systems for Information Flow Control: Granularity and
%  Semantic Models}
\title{Types for Information Flow Control: Labeling Granularity and
  Semantic Models}

\author{
\IEEEauthorblockN{Vineet Rajani}
\IEEEauthorblockA{Max Planck Institute for Software Systems\\Saarland Informatics Campus\\Germany}
\and
\IEEEauthorblockN{Deepak Garg}
\IEEEauthorblockA{Max Planck Institute for Software Systems\\Saarland Informatics Campus\\Germany}
}
\maketitle

% \IEEEpeerreviewmaketitle

\begin{abstract}
Language-based information flow control (IFC) tracks dependencies
within a program using sensitivity labels and prohibits public outputs
from depending on secret inputs. In particular, literature has
proposed several type systems for tracking these dependencies. On one
extreme, there are fine-grained type systems (like Flow Caml) that
label all values individually and track dependence at the level of
individual values. On the other extreme are coarse-grained type
systems (like HLIO) that track dependence coarsely, by associating a
single label with an entire computation context and not labeling all
values individually.

In this paper, we show that, despite their glaring differences, both
these styles are, in fact, equally expressive. To do this, we show a
semantics- and type-preserving translation from a coarse-grained type
system to a fine-grained one and vice-versa. The forward translation
isn't surprising, but the backward translation is: It requires a
construct to arbitrarily limit the scope of a context label in the
coarse-grained type system (e.g., HLIO's ``toLabeled'' construct). As
a separate contribution, we show how to extend work on logical
relation models of IFC types to higher-order state. We build such
logical relations for both the fine-grained type system and the
coarse-grained type system. We use these relations to prove the two
type systems and our translations between them sound.

  %% Information Flow Control (IFC) is a form of dependence analysis that tracks and prohibits
  %% dependence of secret inputs on public outputs. Such dependence analysis is often carried out at
  %% the level of type systems. These IFC type systems can track dependence (via confidentiality
  %% labels) at varying levels of granularity. On one extreme, there are fine-grained type systems
  %% (like FlowCaml) that track dependence at the level of individual values by using labels on their
  %% types. As a result all types in the language are annotated with a label. On the other extreme,
  %% there are coarse-grained type systems (like HLIO) that only track dependence at the level of
  %% entire computation (computation's labels is a lower bound on the level of all the secrets it has
  %% read), represented at the type level using a monad. Since, tracking of labels only happen inside a
  %% monad, there is no need to attach labels on all types.

  %% There has been a growing confusion around the relative expressiveness of these two classes of type
  %% systems. Prior work has made partial progress on this problem by showing a type-preserving
  %% translation from a variant of HLIO to a variant of FlowCaml but not vice versa. In this paper we
  %% come up with a type-preserving translation in the other direction too, prove that both the
  %% translations preserve the semantics and the security of the source program. This resolves the
  %% confusion around their relative expressiveness and shows that the two are equivalent.
\end{abstract}

%%% Local Variables:
%%% mode: latex
%%% TeX-master: "main"
%%% End:

\section{Introduction}
\label{sec:intro}

Information flow control (IFC) is the problem of tracking flows of
information within a computer system and controlling or prohibiting
flows that contravene the policy in effect. In a language-based
setting, IFC requires tracking \emph{dependencies} between a program's
inputs, intermediate values and outputs. This can be done dynamically
with runtime monitoring~\cite{plas09-austinFlanagan,
  plas10-austinFlanagan} or statically using some form of abstraction
interpretation such as a type system. Our focus in this paper is the
second of these methods---IFC enforced through type systems.
In fact, literature has proposed several type systems for IFC, e.g.,
\cite{toplas03-flowcaml,jcs96-volpanoSmith,tosem00-jif,aplas13-paragon,popl98-SLAM,DBLP:journals/jcs/MatosB09,DBLP:conf/Boudol08,icfp15-HLIO,popl99-DCC}. All
these type systems have one aspect in common: \update{They all
  introduce \emph{security labels} or \emph{levels}, elements of a
  security lattice, that abstract program values. These labels are
  used to track dependencies between program values.}

\update{A significant design consideration for an IFC type system is
  the granularity (or extent) of the label abstraction, and the effect
  of this granularity on the expressiveness of the type system. By
  expressiveness here we mean the ability of a type system to type as
  many semantically secure programs as possible.\footnote{No sound
    type system can type all semantically secure programs, since
    freedom from bad flows (specifically, the standard information
    flow security property called noninterference~\cite{goguen82:ni})
    is undecidable.}  More specifically, we call a type system $T$
  more expressive than a type system $T'$ if there is a compositional,
  semantics-preserving transformation of programs typed under $T'$ to
  programs typed under $T$.\footnote{This notion of expressiveness is
    closely related to what Felleisen calls macro
    expressibility~\cite{DBLP:journals/scp/Felleisen91}.}}

\update{
The question of granularity has at least two aspects.
First, one may vary the granularity of the \emph{labels}
themselves. For example, a fine-grained label on a value may precisely
specify the program variables or inputs that the value depends on. On
the other hand, a coarse-grained label may specify only an upper-bound
on the confidentiality level (e.g., ``secret'', ``top-secret'', etc.)
of all inputs on which the labeled value depends. The effect of
varying this notion of granularity (of the labels) on the
expressiveness of the type system has been studied in prior
work~\cite{DBLP:conf/popl/HuntS06}.}

\update{A different kind of granularity, whose expressiveness is the
  focus of this paper, is the granularity of \emph{labeling} (not
  \emph{labels}).%
  \footnote{In the rest of the paper, granularity refers to the
    granularity of labeling, not the granularity of labels.}
  Here,} a \emph{fine-grained type
  system} is one that labels every program value individually. For
instance, Flow Caml, a type system for IFC on
ML~\cite{toplas03-flowcaml}, adds a label on every type constructor
and, hence, on every value, top-level and nested. As an example, the
type $(A^H \times B^L)^L$ might ascribe low (public) pairs, whose
first projection is high (private) and whose second projection is
low. ($H$ and $L$ are standard labels for high and low confidentiality
data, respectively.) Since fine-grained type systems label individual
values, they also track dependencies at the granularity of individual
values. For example, combining a high value with a low value using a
primitive operator in the language results in a high value. Many other
type systems are similarly
fine-grained~\cite{jcs96-volpanoSmith,tosem00-jif,aplas13-paragon,popl98-SLAM}.

In contrast, a \emph{coarse-grained type system} labels an entire
sub-computation using a single label. All values produced within the
scope of the sub-computation implicitly have that label. Hence, it not
necessary to label individual values.
As an example, the SLIO and HLIO systems~\cite{icfp15-HLIO} introduce
a monad for heap I/O (similar to Haskell's IO monad), but refine the
monadic type to include two labels, $\llabel_i$ and $\llabel_o$, as in
$(SLIO~\llabel_i~\llabel_o~\tau)$. This type represents stateful
computations of type $\tau$ that start from the confidentiality label
$\llabel_i$ and end with the confidentiality label
$\llabel_o$. $\llabel_i$ is an upper-bound on the confidentiality of
all prior computations that the current computation depends on;
accordingly, the current computation can have write effects at levels
above $\llabel_i$ only. $\llabel_o$ is an upper-bound on the
confidentiality of the current computation; accordingly, the current
computation can have read effects only at levels below
$\llabel_o$. Importantly, there is no need to label individual values
or nested types. Instead, every value produced by the current
computation implicitly inherits the label $\llabel_o$, and labels are
tracked via monadic sequencing (bind) at the granularity of
computations. Other type
systems~\cite{DBLP:journals/jcs/MatosB09,DBLP:conf/Boudol08,popl99-DCC}
are also coarse-grained,
although~\cite{DBLP:journals/jcs/MatosB09,DBLP:conf/Boudol08} do not
use monads to confine effects.

Given these vastly contrasting labeling granularities for the same
end-goal---information flow control, a natural question is one of
their relative expressiveness~\cite{siglog17-ifcComp}.
%% By expressiveness here we mean the ability of a type system to type
%% as many semantically secure programs as possible.\footnote{No sound
%% type system can type all semantically secure programs, since
%% freedom from bad flows (specifically, the standard security
%% property called noninterference) is undecidable.}
In general, it seems that fine-grained type systems should be at least
as expressive as coarse-grained type systems, since the former track
flows at finer granularity \update{(individual values as opposed to
  entire sub-computations)} and, hence, should abstract flows less
than the latter. In the other direction, the situation is less
clear. Upfront, it seems that coarse-grained type systems \emph{may}
be less expressive than fine-grained type systems, but then one
wonders whether by structuring programs as extremely small
computations in a coarse-grained type system, one may recover the
expressiveness of a fine-grained type system.

In this paper, we show constructively that both these intuitions are,
in fact, correct. We do this using specific instances of the two kinds
of type systems in the setting of a higher-order language with state
(similar to ML). For the fine-grained type system we use a system very
close to SLam~\cite{popl98-SLAM} and the exception-free fragment of
Flow Caml~\cite{toplas03-flowcaml}. For the coarse-grained type system
we use a variant of the static fragment of
HLIO~\cite{icfp15-HLIO}. This calculus has a specific construct to
limit the scope of a computation's label in a safe way. We then show
that well-typed programs in each type system can be translated to the
other, preserving typability and meaning. This establishes that the
type systems are equally expressive.

We believe this settles an open question about the relative
expressiveness of setting up IFC type systems \update{with different
  labeling granularities}. Our result also has an immediate practical
consequence: Since coarse-grained IFC type systems usually burden a
programmer less with annotations (since not every value has to be
labeled) and we have shown now that they are as expressive as
fine-grained IFC type systems, there seems to be some merit to
preferring coarse-grained IFC type systems over fine-grained ones in
general.

As a second contribution of independent interest, we show how to set
up semantic, logical relations models of IFC types in both the
fine-grained and the coarse-grained settings, over calculi with
higher-order state. While models of IFC types have been considered
before~\cite{popl98-SLAM,popl99-DCC,esop99-PER-IFC,DBLP:conf/csfw/MantelSS11},
we do not know of any development that covers higher-order state.  In
fact, models of types in the presence of higher-order state are
notoriously difficult. Here, we have the added complication of
information flow labels. Fortunately, enough development has occurred
in the programming languages community in the past decade to give us a
good starting point. Specifically, our models are based on
step-indexed Kripke logical
relations~\cite{DBLP:conf/popl/AhmedDR09}. Like earlier work, our
models are relational, i.e., they relate two runs of a program to each
other. This is essential since we are interested in proving
noninterference~\cite{goguen82:ni}, the standard security property
which says that public outputs of a program are not influenced by
private inputs (i.e., there are no bad flows). This property is
naturally defined using two runs. Using our models, we derive proofs
of the soundness of both the fine-grained and the coarse-grained type
systems.

%% Prior proofs of the soundness of similar systems are either fully
%% syntactic (based directly on the operational semantics) and far
%% more complex~\cite{toplas03-flowcaml,icfp15-HLIO} or ignore the
%% stateful part~\cite{popl98-SLAM}.

We also use our logical relations to show that our translations are
meaningful. Specifically, we set up \emph{cross-language logical
  relations} to prove that our translations preserve program
semantics, and from this, we derive a crucial result for each
translation: Using the noninterference theorem of the target language
as a lemma, we are able to re-prove the noninterference theorem for
the source language directly. These results imply that our
translations preserve label annotations
meaningfully~\cite{DBLP:journals/cl/BartheRB07}. Like all logical
relations models, we expect that our models can be used for other
purposes as well.

To summarize, the two contributions of this work are:
\begin{itemize}
\item Typability- and meaning-preserving translations between a
  fine-grained and a coarse-grained IFC type system, showing that
  these type systems are equally expressive.
\item Logical relations models of both type systems, covering both
  higher-order functions and higher-order state.
\end{itemize}

Due to lack of space, many technical details and proofs are omitted
from this paper. These are provided in an appendix available online
from the authors' homepages.

\medskip
\noindent \textbf{Note.} Readers interested only in our translations
but not the details of our semantic models can skip sections
pertaining to the latter (e.g., Section~\ref{sec:fg-sem}). This will
not affect the readability of the rest of the paper.

\section{The Two Type Systems}
\label{sec:ts}

In this section, we describe the fine-grained and coarse-grained type
systems we work with. Both type systems are set up for higher-order
stateful languages, but differ considerably in how they enforce
IFC. The fine-grained type system, called {\fg}, works on a language
with pervasive side-effects like ML, and associates a security label
with every expression in the language. The coarse-grained type system,
{\cg}, works on a language that isolates state in a monad (like
Haskell's IO monad) and tracks flows coarsely at the granularity of a
monadic computation, not on pure values within a monadic computation.

%% In this section we describe the two type systems that we work with, namely {\fg} and {\cg}. Both the
%% type systems are setup for higher order stateful languages, but they differ considerably in the way
%% they enforce IFC. {\fg} is for a language with pervasive side effects, like ML, and tracks security
%% labels with every value in the language. In contrast, {\cg} is for a language that isolates the
%% side-effects in a monad, like Haskell, and tracks flows only within a monad and not across pure
%% expressions.

Both {\fg} and {\cg} use security labels (denoted by $\llabel$) drawn
from an arbitrary security lattice ($\lattice, \lbelow$). We denote
the least and top element of the lattice by $\bot$ and $\top$
respectively. As usual, the goal of the type systems is to ensure that
outputs labeled $\llabel$ depend only on inputs with security labels
$\llabel$ or lower. \update{For drawing intuitions, we find it
  convenient to think of a confidentiality lattice (labels higher in
  the lattice represent higher confidentiality). However, nothing in
  our technical development is specific to a confidentiality
  lattice---the development works for any security lattice including
  an integrity lattice and a product lattice for confidentiality and
  integrity.}

\subsection{The fine-grained type system, {\fg}}
\label{sec:fg}

\begin{figure*}[!htbp]
\newcolumntype{L}{>{$}l<{$}}
\newcolumntype{P}{>{$}X<{$}}

\begin{tabularx}{\textwidth}{lLLP}
  Expressions & e & ::= & x \mid \lambda x.e \mid e \; e \mid (e, e) \mid \efst(e) \mid \esnd(e)
  \mid \einl(e) \mid \einr(e) \mid \ecase(e, x.e, x.e) \mid \enew e \mid {!}e \mid {e := e} \\

%  Labels & \llabel, \pc & ::= & \bot \mid \top \mid \lelement \mid \llabel \ljoin \llabel \mid \llabel \lmeet \llabel\\

  (Labeled) Types & \tau & ::= & \utype^\llabel \\

  Unlabeled types & \utype & ::= & \tbase \mid \tunit \mid \tau
  \mathbin{\smash{\overset{\llabel_e}{\to}}} \tau \mid \tau \times \tau \mid \tau + \tau \mid\tref ~
  \tau ~~~~ (\tbase \mbox{ denotes a base type})
\end{tabularx}\\ \\

\textbf{Typing judgment:} \framebox{$\Gamma \vdash_\pc e: \tau$}

\begin{mathpar}
  \inferrule
    { }
    {
      \Gamma, x:\tau \vdash_\pc x: \tau
    }\rname{{\fg}-var}\vspace{-.2em}

    \inferrule
    {
      \Gamma, x:\tau_1 \vdash_{\llabel_e} e: \tau_2
    }
    {
      \Gamma \vdash_\pc \lambda x. e: (\tau_1 \fto \tau_2)^\bot
    }\rname{{\fg}-lam}\vspace{-.2em}

    \inferrule
    {
      \Gamma \vdash_\pc e_1: (\tau_1 \fto \tau_2)^\llabel
      \\ \Gamma \vdash_\pc e_2: \tau_1
      \\ \lattice \vdash \tau_2 \searrow \llabel
      \\ \lattice \vdash \pc \ljoin \llabel \lbelow \llabel_e
    }
    {
      \Gamma \vdash_\pc e_1 ~ e_2: \tau_2
    }\rname{{\fg}-app}\vspace{-.2em}

    \inferrule
    {
      \Gamma \vdash_\pc e_1: \tau_1
      \\ \Gamma \vdash_\pc e_2: \tau_2
    }
    {
      \Gamma \vdash_\pc (e_1,e_2): (\tau_1 \times \tau_2)^{\bot}
    }\rname{{\fg}-prod}\vspace{-.2em}

    \inferrule
    {
      \Gamma \vdash_\pc e: (\tau_1 \times \tau_2)^{\llabel}
      \\ \lattice \vdash \tau_1 \searrow \llabel
    }
    {
      \Gamma \vdash_\pc \efst (e): \tau_1
    }\rname{{\fg}-fst}\vspace{-.2em}

    \quad

    \inferrule
    {
      \Gamma \vdash_\pc e: \tau_1
    }
    {
      \Gamma \vdash_\pc \einl (e): (\tau_1 + \tau_2)^\bot
    }\rname{{\fg}-inl}\vspace{-.2em}

    \inferrule
    {
      \Gamma \vdash_\pc e: (\tau_1 + \tau_2)^\llabel
      \\ \Gamma, x:\tau_1 \vdash_{\pc \ljoin \llabel} e_1: \tau
      \\ \Gamma, y:\tau_2 \vdash_{\pc \ljoin \llabel} e_2: \tau
      \\ \lattice \vdash \tau \searrow \llabel
    }
    {
      \Gamma \vdash_\pc \ecase (e, x.e_1, y.e_2): \tau
    }\rname{{\fg}-case}\vspace{-.2em}

    \inferrule
    {
      \Gamma \vdash_{\pc'} e: \tau'
      \\ \lattice \vdash \pc \lbelow \pc'
      \\ \lattice \vdash \tau' \subtype \tau
    }
    {
      \Gamma \vdash_\pc e: \tau
    }\rname{{\fg}-sub}\vspace{-.2em}

    \inferrule
    {
      \Gamma \vdash_\pc e : \tau
      \\ \lattice \vdash \tau \searrow \pc
    }
    {
      \Gamma \vdash_\pc \enew e : (\tref~ \tau)^\bot
    }\rname{{\fg}-ref}\vspace{-.2em}

    \inferrule
    {
      \Gamma \vdash_\pc e : (\tref~ \tau)^\llabel
      \\ \lattice \vdash \tau \subtype \tau'
      \\ \lattice \vdash \tau' \searrow \llabel
    }
    {
      \Gamma \vdash_\pc {!}e : \tau'
    }\rname{{\fg}-deref}\vspace{-.2em}

    \inferrule
    {
      \Gamma \vdash_\pc e_1 : (\tref~ \tau)^\llabel
      \\ \Gamma \vdash_\pc e_2 : \tau
      \\ \lattice \vdash \tau \searrow (\pc \ljoin \llabel)
    }
    {
      \Gamma \vdash_\pc e_1 := e_2 : \tunit
    }\rname{{\fg}-assign}\vspace{-.2em}

    \inferrule
    { }
    {\Gamma \vdash_\pc (): \tunit^\bot}
    \rname{{\fg}-unitI}\vspace{-.2em}
  \end{mathpar} \\

  \textbf{Subtyping judgments:} \framebox{$\lattice \vdash \utype \subtype \utype'$} and
  \framebox{$\lattice \vdash \tau \subtype \tau'$}

  \begin{mathpar}
    \inferrule
    {\lattice \vdash \llabel \lbelow \llabel' \\
      \lattice \vdash \utype \subtype \utype'
    }
    {
      \lattice \vdash \utype^\llabel \subtype \utype'^{\llabel'}
    }\rname{{\fg}sub-label}\vspace{-.2em}

    %% \inferrule
  %% {  }
  %% { \lattice \vdash \tbase \subtype \tbase }\rname{{\fg}sub-base}

  \inferrule
  { }
  {
    \lattice \vdash \tref~ \tau \subtype \tref~ \tau
  }\rname{{\fg}sub-ref}\vspace{-.2em}

  %% \inferrule
  %% { \lattice \vdash \tau_1 \subtype \tau_1'
  %%   \\ \lattice \vdash \tau_2 \subtype \tau_2'}
  %% { \lattice \vdash \tau_1 \times \tau_2
  %%   \subtype \tau_1' \times \tau_2' }\rname{{\fg}sub-prod}

  %% \inferrule
  %% { \lattice \vdash \tau_1 \subtype \tau_1'
  %%   \\ \lattice \vdash \tau_2 \subtype \tau_2'}
  %% { \lattice \vdash \tau_1 + \tau_2
  %%   \subtype \tau_1' + \tau_2' }\rname{{\fg}sub-sum}

  \inferrule
  { \lattice \vdash \tau_1' \subtype \tau_1
    \\ \lattice \vdash \tau_2 \subtype \tau_2'
    \\ \lattice \vdash \llabel_e' \lbelow \llabel_e }
  {
    \lattice \vdash \tau_1 \fto \tau_2 \subtype
    \tau_1' \ftoP \tau_2'
  }\rname{{\fg}sub-arrow}\vspace{-.2em}

  %% \inferrule
  %% { }
  %% {\lattice \vdash \tunit \subtype \tunit}\rname{{\fg}sub-unit}
\end{mathpar}

\caption{{\fg}'s language syntax and type system (selected rules)}
\label{fig:fg-ts}
\end{figure*}

%% \begin{figure*}[!htbp]
%%   \centering
%% \begin{mathpar}

%%   \inferrule
%%   {\lattice \vdash \llabel \lbelow \llabel' \\
%%     \lattice \vdash \utype \subtype \utype'
%%   }
%%   {\lattice \vdash \utype^\llabel \subtype \utype'^{\llabel'}}\rname{{\fg}sub-label}

%%   \inferrule
%%   {  }
%%   { \lattice \vdash \tbase \subtype \tbase }\rname{{\fg}sub-base}

%%   \inferrule
%%   { }
%%   { \lattice \vdash \tref~ \tau \subtype \tref~ \tau }\rname{{\fg}sub-ref}

%%   \inferrule
%%   { \lattice \vdash \tau_1 \subtype \tau_1'
%%     \\ \lattice \vdash \tau_2 \subtype \tau_2'}
%%   { \lattice \vdash \tau_1 \times \tau_2
%%     \subtype \tau_1' \times \tau_2' }\rname{{\fg}sub-prod}

%%   \inferrule
%%   { \lattice \vdash \tau_1 \subtype \tau_1'
%%     \\ \lattice \vdash \tau_2 \subtype \tau_2'}
%%   { \lattice \vdash \tau_1 + \tau_2
%%     \subtype \tau_1' + \tau_2' }\rname{{\fg}sub-sum}

%%   \inferrule
%%   { \lattice \vdash \tau_1' \subtype \tau_1
%%     \\ \lattice \vdash \tau_2 \subtype \tau_2'
%%     \\ \lattice \vdash \llabel_e' \lbelow \llabel_e }
%%   { \lattice \vdash \tau_1 \fto \tau_2 \subtype
%%     \tau_1' \ftoP \tau_2' }\rname{{\fg}sub-arrow}

%%   \inferrule
%%   { }
%%   {\lattice \vdash \tunit \subtype \tunit}\rname{{\fg}sub-unit}
%% \end{mathpar}

%% \caption{{\fg} subtyping}
%% \label{fig:fg-sub}
%% \end{figure*}
%%% Local Variables:
%%% mode: latex
%%% TeX-master: "main"
%%% End:

{\fg} is based on the SLam calculus~\cite{popl98-SLAM}, but uses a
presentation similar to Flow Caml, an IFC type system for
ML~\cite{toplas03-flowcaml}. It works on a call-by-value, eager
language, which is a simplification of ML. The syntax of the language
is shown at the top of Figure~\ref{fig:fg-ts}. The language has all
the usual expected constructs: Functions, pairs, sums, and mutable
references (heap locations). The expression ${!}e$ dereferences the
location that $e$ evaluates to, while ${e_1 := e_2}$ assigns the value
that $e_2$ evaluates to, to the location that $e_1$ evaluates to. The
dynamic semantics of the language are defined by a ``big-step''
judgment $(\heap, e) \ReducesTo{j} (\heap', \val)$, which means that
starting from heap $\heap$, expression $e$ evaluates to value $\val$,
ending with heap $H'$. This evaluation takes $j$ steps. The number of
steps is important only for our logical relations models. The rules
for the big-step judgment are standard, hence omitted here.

Every type $\tau$ in {\fg}, including a type nested inside another,
carries a security label. The security label represents the
confidentiality level of the values the type ascribes. It is also
convenient to define unlabeled types, denoted $\utype$, as shown in
Figure~\ref{fig:fg-ts}.

\medskip
\noindent\textbf{Typing rules.}  {\fg} uses the typing judgment
$\Gamma \vdash_\pc e: \tau$. As usual, $\Gamma$ maps free variables of
$e$ to their types. The judgment means that, given the types for free
variables as in $\Gamma$, $e$ has type $\tau$. The annotation $\pc$ is
also a label drawn from $\lattice$, often called the ``program
counter'' label. This label is a \emph{lower bound} on the write
effects of $e$. The type system ensures that any reference that $e$
writes to is at a level $\pc$ or higher. This is necessary to prevent
information leaks via the heap. A similar annotation, $\llabel_e$,
appears in the function type $\tau_1 \fto \tau_2$. Here, $\llabel_e$
is a lower bound on the write effects of the body of the function.

{\fg}'s typing rules are shown in Figure~\ref{fig:fg-ts}. We describe
some of the important rules. In the rule for case analysis
({{\fg}-case}), if the case analyzed expression $e$ has label
$\llabel$, then both the case branches are typed in a $\pc$ that is
\emph{joined} with $\llabel$. This ensures that the branches do not
have write effects below $\llabel$, which is necessary for IFC since
the execution of the branches is control dependent on a value (the
case condition) of confidentiality $\llabel$. Similarly, the type of
the result of the case branches, $\tau$, must have a top-level label
at least $\llabel$. This is indicated by the premise $\tau \searrow
\llabel$ and prevents implicit leaks via the result. The relation
$\tau \searrow \llabel$, read ``$\tau$ protected at
$\llabel$''~\cite{popl99-DCC}, means that if $\tau =
\utype^{\llabel'}$, then $\llabel \lbelow \llabel'$.

The rule for function application ({{\fg}-app}) follows similar
principles. If the function expression $e_1$ being applied has type
$(\tau_1 \fto \tau_2)^{\llabel}$, then $\llabel$ must be below
$\llabel_e$ and the result $\tau_2$ must be protected at $\llabel$ to
prevent implicit leaks arising from the identity of the function that
$e_1$ evaluates to.

In the rule for assignment ({{\fg}-assign}), if the expression
$e_1$ being assigned has type $(\tref~ \tau)^\llabel$, then $\tau$
must be protected at $\pc \ljoin \llabel$ to ensure that the written
value (of type $\tau$) has a label above $\pc$ and $\llabel$. The
former enforces the meaning of the judgment's $\pc$, while the latter
protects the identity of the reference that $e_1$ evaluates to.

All introduction rules such as those for $\lambda$s, pairs and sums
produce expressions labeled $\bot$. This label can be weakened
(increased) freely with the subtyping rule {{\fg}sub-label}. The
other subtyping rules are the expected ones, e.g., subtyping for
unlabeled function types $\tau_1 \fto \tau_2$ is co-variant in
$\tau_2$ and contra-variant in $\tau_1$ and $\llabel_e$
(contra-variance in $\llabel_e$ is required since $\llabel_e$ is a
\emph{lower} bound on an effect). Subtyping for $\tref~\tau$ is
invariant in $\tau$, as usual.

The main meta-theorem of interest to us is soundness. This theorem
says that every well-typed expression is \emph{noninterferent}, i.e.,
the result of running an expression of a type labeled low is
independent of substitutions used for its high-labeled free
variables. This theorem is formalized below. Note that we work here
with what is called termination-insensitive noninterference; we
briefly discuss the termination-sensitive variant in
Section~\ref{sec:discussion}.
\begin{thm}[Noninterference for {\fg}]
    \label{thm:ni-fg}
  Suppose (1) $\llabel_i \not\lbelow \llabel$, (2)
  $x : \utype^{\llabel_i} \vdash_\pc e: \tbool^\llabel$, and (3)
  $ v_1, v_2: \smash{\smash{\utype}^{\smash{\llabel}_i}}$. If both $e[v_1/x]$ and $e[v_2/x]$
  terminate, then they produce the same value (of type $\tbool$).
\end{thm}

By definition, noninterference, as stated above is a relational
(binary) property, i.e., it relates two runs of a program. Next, we
show how to build a semantic model of {\fg}'s types that allows
proving this property.

\subsubsection{Semantic model of {\fg}}\label{sec:fg-sem}
\begin{figure*}[!htbp]
\begin{displaymath}
  \begin{array}{lll}
    \Ulrv {\tbase} & \triangleq & \{(\uWorld, m, \val) ~|~ \val \in \llbracket \tbase \rrbracket \} \\

    \Ulrv {\tunit} & \triangleq & \{(\uWorld, m,  \val) ~|~ \val \in \llbracket \tunit \rrbracket \} \\

    \Ulrv {\tau_1 \times \tau_2} & \triangleq & \{(\uWorld, m, (\val_1, \val_2)) ~|~
                                       (\uWorld, m, \val_1) \in \Ulrv {\tau_1} \wedge
                                       (\uWorld, m, \val_2) \in \Ulrv {\tau_2}
                                       \}\\

    \Ulrv {\tau_1 + \tau_2} & \triangleq & \{(\uWorld, m, \einl ~  \val) ~|~
                                           (\uWorld, m, \val) \in \Ulrv {\tau_1} \} \union
                                           \{(\uWorld, m, \einr ~  \val) ~|~
                                           (\uWorld, m, \val) \in \Ulrv {\tau_2} \} \\

    \Ulrv {\tau_1 \fto \tau_2} & \triangleq & \{(\uWorld, m, \lambda x. e) ~|~
                                              \forall \uWorld'. \uWorld \wExtends \uWorld' \wedge \forall j < m.
                                              \forall \val. ((\uWorld', j, \val) \in \Ulrv {\tau_1} \implies
                          (\uWorld', j, e[\val/x]) \in \lreu {\tau_2} {\llabel_e}) \} \\

    \Ulrv {\tref ~ \tau} & \triangleq & \{(\uWorld, m, \loc) ~|~  \uWorld (\loc) = \tau
                                        \} \\ \\
%%%%%%%%%%%%%%%%%%%%%%%%%%%%%%%%%%%%%%%%%%%%%%%%%%%%%%%
    \Ulrv {\utype^{\llabel}} & \triangleq & \Ulrv {\utype}
    \\ \\
%%%%%%%%%%%%%%%%%%%%%%%%%%%%%%%%%%%%%%%%%%%%%%%%%%%%%%%
\hline \\
    \Ulre {\tau} & \triangleq & \{(\uWorld, n, e) ~|~
                                \forall \heap. (n, \heap) \wDefined {} \uWorld \wedge
                                \forall j < n. (\heap, e) \ReducesTo{j} (\heap', \val') \implies \\

                 &    & ~~\exists \uWorld'. \uWorld \wExtends \uWorld' \wedge
                        (n-j, \heap') \wDefined{} \uWorld' \wedge
                        (\uWorld', n-j, \val') \in \Ulrv {\tau} \wedge \\
                 &   &  ~~~~~~(\forall \loc. \heap(\loc) \neq \heap'(\loc) \implies
                        \exists \llabel'. \uWorld(\loc) = \utype^{\llabel'} \wedge \pc \lbelow \llabel'
                        ) \wedge\\
                 &    &  ~~~~~~ (\forall \loc \in \dom(\uWorld') \backslash \dom(\uWorld). \uWorld'(\loc) \searrow \pc)
                        \}\\ \\
%%%%%%%%%%%%%%%%%%%%%%%%%%%%%%%%%%%%%%%%%%%%%%%%%%%%%%%%
\hline \\
    (n,\heap) \wDefined{} \uWorld & \triangleq & \dom(\uWorld) \subseteq \dom(\heap) \wedge
                                            \forall \loc \in \dom(\uWorld). (\uWorld, n-1, \heap(\loc)) \in
                                            \Ulrv {\uWorld(\loc)}
  \end{array}
\end{displaymath}
\caption{Unary value, expression, and heap conformance relations for {\fg}}
\label{fig:ulr-fg}
\end{figure*}

\begin{figure*}[!htbp]
\begin{displaymath}
  \begin{array}{lll}
    \Blrv {\tbase} & \triangleq & \{(\world, n, \val_1, \val_2) ~|~ \val_1 = \val_2 \wedge \{\val_1,\val_2\}
                                  \in \llbracket \tbase \rrbracket \} \\

    \Blrv {\tunit} & \triangleq & \{(\world, n, (), ()) ~|~  () \in \llbracket \tunit \rrbracket \} \\

    \Blrv {\tau_1 \times \tau_2} & \triangleq & \{(\world, n, (\val_1, \val_2), (\val_1', \val_2')) ~|~
                                                (\world, n, \val_1, \val_1') \in \Blrv {\tau_1} \wedge
                                                (\world, n, \val_2, \val_2') \in \Blrv {\tau_2}
                                                \} \\

    \Blrv {\tau_1 + \tau_2} & \triangleq & \{(\world, n, \einl ~  \val, \einl ~  \val') ~|~
                                           (\world, n, \val, \val') \in \Blrv {\tau_1}
                                           \} \union\\
                   & & \{(\world, n, \einr ~  \val, \einr ~  \val') ~|~
                       (\world, n, \val, \val') \in \Blrv {\tau_2}
                       \} \\

    \Blrv {\tau_1 \fto \tau_2} & \triangleq & \{(\world, n, \lambda x. e_1, \lambda x. e_2) ~|~\\
                   & & ~~\forall \world' \wExtendsR \world, j < n, \val_1, \val_2.\\
                   & & ~~~~~~~~~((\world', j, \val_1, \val_2) \in \Blrv {\tau_1}
                       \implies (\world', j, e_1[\val_1/x], e_2[\val_2/x]) \in \Blre {\tau_2} ) \wedge \\
                   & & ~~\forall \uWorld_l \wExtendsR \world.\uWorld_1, j, \val_c.
                   \, ((\uWorld_l, j, \val_c) \in \Ulrv {\tau_1} \implies
                       (\uWorld_l, j, e_1[\val_c/x]) \in \lreu {\tau_2} {\llabel_e}) \wedge  \\
                   & & ~~\forall \uWorld_l \wExtendsR \world.\uWorld_2, j, \val_c.
                   \, ((\uWorld_l, j, \val_c) \in \Ulrv {\tau_1} \implies
                       (\uWorld_l, j, e_2[\val_c/x]) \in \lreu {\tau_2} {\llabel_e}) \} \\

    \Blrv {\tref ~ \tau} & \triangleq & \{(\world, n, \loc_1,  \loc_2) ~|~
                     (\loc_1 , \loc_2) \in \world.\pb
                       \wedge \world.\uWorld_1(\loc_1) = \world.\uWorld_2(\loc_2) = \tau \}\\ \\
%%%%%%%%%%%%%%%%%%%%%%%%%%%%%%%%%%%%%%%%%%%%%%%%%%%%%%%%%%%%%%%%
\Blrv {\utype^{\llabel}} & \triangleq &
\left\lbrace
  \begin{array}{l r r}
    \{(\world, n, \val_1,  \val_2) ~|~
    (\world, n, \val_1, \val_2) \in \Blrv {\utype} \} & & \llabel \lbelow \attacker \\

    \{(\world, n, \val_1,  \val_2) ~|~
    \forall i\in \{1,2\}. \forall m. (\world.\uWorld_i, m, \val_i) \in \Ulrv {\utype} \} & & \llabel \not\lbelow \attacker
  \end{array}
\right. \\ \\
%%%%%%%%%%%%%%%%%%%%%%%%%%%%%%%%%%%%%%%%%%%%%%%%%%%%%%%%%%%%%%%%
\hline \\
    \Blre {\tau} & \triangleq & \{(\world, n, e_1, e_2) ~|~\\
                   &   & ~~\forall \heap_1, \heap_2, j<n. (n, \heap_1, \heap_2) \wDefined {\attacker} \world \wedge
                   (\heap_1, e_1) \ReducesTo{j} (\heap_1', \val_1') \wedge
                         (\heap_2, e_2) \ReducesTo{} (\heap_2', \val_2') \implies \\

                   &   & ~~~~~~~\exists \world' \wExtendsR \world. (n - j, \heap_1', \heap_2') \wDefined {\attacker} \world' \wedge
                         (\world', n - j, \val_1', \val_2') \in \Blrv {\tau} \}
    \\ \\
%%%%%%%%%%%%%%%%%%%%%%%%%%%%%%%%%%%%%%%%%%%%%%%%%%%%%%%%%%%%%%%%
    \hline \\
    (n, \heap_1, \heap_2) \wDefined{\attacker} \world & \triangleq & \dom(\world.\uWorld_1) \subseteq \dom(\heap_1) \wedge
                                                                   \dom(\world.\uWorld_2) \subseteq \dom(\heap_2) \wedge
                                                     (\world.\pb) \subseteq (\dom(\world.\uWorld_1) \times \dom(\world.\uWorld_2))
                                                           \wedge \\
                                                    &    & \forall (\loc_1,\loc_2) \in (\world.\pb).
                                                           (\world.\uWorld_1(\loc_1) = \world.\uWorld_2(\loc_2) \wedge
                                                     (\world, n-1, \heap_1(\loc_1),\heap_2(\loc_2)) \in
                                                           \Blrv {\world.\uWorld_1(\loc_1)})
                                                           \wedge \\
                                                    &   & \forall i \in \{1,2\}. \forall m. \forall \loc_i \in \dom(\world.\uWorld_i).
                                                          (\world.\uWorld_i, m, \heap_i(\loc_i)) \in \Ulrv {\world.\uWorld_i(\loc_i)}
  \end{array}
\end{displaymath}

\caption{Binary value, expression and heap conformance relations for {\fg}}
\label{fig:blr-fg}
\end{figure*}

%%% Local Variables:
%%% mode: latex
%%% TeX-master: "main"
%%% End:

We now describe our semantic model of {\fg}'s types. We use this model
to show that the type system is sound (Theorem~\ref{thm:ni-fg}) and
later to prove the soundness of our translations. Our semantic model
uses the technique of step-indexed Kripke logical
relations~\cite{DBLP:conf/popl/AhmedDR09} and is more directly based
on a model of types in a different domain, namely, incremental
computational complexity~\cite{DBLP:conf/icfp/CicekP016}. In
particular, our model captures all the invariants necessary to prove
noninterference.

The central idea behind our model is to interpret each type in two
different ways---as a set of values (unary interpretation), and as a
set of pairs of values (binary interpretation). The binary
interpretation is used to relate \emph{low}-labeled types in the two
runs mentioned in the noninterference theorem, while the unary
interpretation is used to interpret \emph{high}-labeled types
independently in the two runs (since high-labeled values may be
unrelated across the two runs). What is high and what is low is
determined by the level of the observer (adversary), which is a
parameter to our binary interpretation.

\textit{Remark.} Readers familiar with earlier models of IFC type
systems~\cite{popl98-SLAM,popl99-DCC,esop99-PER-IFC} may wonder why we
need a unary relation, when prior work did not. The reason is that we
handle an effect (mutable state) in our model, which prior work did
not. In the absence of effects, the unary model is unnecessary. In the
presence of effects, the unary relation captures what is often called
the ``confinement lemma'' in proofs of noninterference---we need to
know that while the two runs are executing high branches
independently, neither will modify low-labeled locations.

\medskip
\noindent \textbf{Unary interpretation.}
The unary interpretation of types is shown in
Figure~\ref{fig:ulr-fg}. The interpretation is actually a Kripke
model. It uses \emph{worlds}, written $\uWorld$, which specify the
type for each valid (allocated) location in the heap. For example,
$\uWorld(a) = \tbool^H$ means that location $a$ should hold a high
boolean. The world can grow as the program executes and allocates more
locations. A second important component used in the interpretation is
a \emph{step-index}, written $m$
or~$n$~\cite{esop06-SILR}. Step-indices are natural numbers, and are
merely a technical device to break a non-well-foundedness issue in
Kripke models of higher-order state, like this one. Our use of
step-indices is standard and readers may ignore them.

The interpretation itself consists of three mutually inductive
relations---a \emph{value relation} for types (labeled and unlabeled),
written $\Ulrv {\tau}$; an \emph{expression relation} for labeled
types, written $\Ulre {\tau}$; and a \emph{heap conformance relation},
written $(n,\heap) \wDefined{} \uWorld$. These relations are
well-founded by induction on the step indices $n$ and types. This is
the only role of step-indices in our model.

The value relation $\Ulrv {\tau}$ defines, for each type, which values
(at which worlds and step-indices) lie in that type. For base types
$\tbase$, this is straightforward: All syntactic values of type
$\tbase$ (written $\llbracket \tbase \rrbracket$) lie in $\Ulrv
{\tbase}$ at any world and any step index. For pairs, the relation is
the intuitive one: $(v_1,v_2)$ is in $\Ulrv{\tau_1 \times \tau_2}$ iff
$v_1$ is in $\Ulrv{\tau_1}$ and $v_2$ is in $\Ulrv{\tau_2}$. The
function type $\tau_1 \fto \tau_2$ contains a value $\lambda x.e$ at
world $\theta$ if in any world $\theta'$ that extends $\theta$, if $v$
is in $\Ulrv{\tau_1}$, then $(\lambda x.e)\ v$ or, equivalently,
$e[v/x]$, is in the \emph{expression relation} $\lreu {\tau_2}
{\llabel_e}$. We describe this expression relation below. Importantly,
we allow for the world $\theta$ to be extended to $\theta'$ since
between the time that the function $\lambda x.e$ was created and the
time that the function is applied, new locations could be
allocated. The type $\tref~\tau$ contains all locations $a$ whose type
according to the world $\theta$ matches $\tau$. Finally, security
labels play no role in the unary interpretation, so $\Ulrv
{\utype^{\llabel}} = \Ulrv {\utype}$ (in contrast, labels play a
significant role in the binary interpretation).

The expression relation $\Ulre {\tau}$ defines, for each type, which
expressions lie in the type (at each $\pc$, each world $\uWorld$ and
each step index $n$). The definition may look complex, but is
relatively straightforward: $e$ is in $\Ulre {\tau}$ if for any heap
$\heap$ that conforms to the world $\theta$ such that running $e$
starting from $\heap$ results in a value $\val'$ and a heap $\heap'$,
there is a some extension $\theta'$ of $\theta$ to which $\heap'$
conforms and at which $v'$ is in $\Ulrv{\tau}$. Additionally, all
writes performed during the execution (defined as the locations at
which $\heap$ and $\heap'$ differ) must have labels above the program
counter, $\pc$. In simpler words, the definition simply says that $e$
lies in $\Ulre {\tau}$ if its resulting value is in $\Ulrv{\tau}$, it
preserves heap conformance with worlds and, importantly, its write
effects are at labels above $\pc$. (Readers familiar with proofs of
noninterference should note that the condition on write effects is our
model's analogue of the so-called ``confinement lemma''.)

The heap conformance relation $(n,\heap) \wDefined{} \uWorld$ defines
when a heap $\heap$ conforms to a world $\uWorld$. The relation is
simple; it holds when the heap $\heap$ maps every location to a value
in the semantic interpretation of the location's type given by the
world $\uWorld$.

\medskip
\noindent \textbf{Binary interpretation.}
The binary interpretation of types is shown in
Figure~\ref{fig:blr-fg}. This interpretation relates two executions of
a program with different inputs. Like the unary interpretation, this
interpretation is also a Kripke model. Its worlds, written $\world$,
are different, though. Each world is a triple $\world = (\uWorld_1,
\uWorld_2, \pb)$. $\uWorld_1$ and $\uWorld_2$ are unary worlds that
specify the types of locations allocated in the two executions. Since
executions may proceed in sync on the two sides for a while, then
diverge in a high-labeled branch, then possibly re-synchronize, and so
on, some locations allocated on one side may have analogues on the
other side, while other locations may be unique to either side. This
is captured by $\pb$, which is a \emph{partial bijection} between the
domains of $\uWorld_1$ and $\uWorld_2$. If $(a_1, a_2) \in \pb$, then
location $a_1$ in the first run corresponds to location $a_2$ in the
second run. Any location not in $\pb$ has no analogue on the other
side.

As before, the interpretation itself consists of three mutually
inductive relations---a \emph{value relation} for types (labeled and
unlabeled), written $\Blrv {\tau}$; an \emph{expression relation} for
labeled types, written $\Blre {\tau}$; and a \emph{heap conformance
  relation}, written $(n, \heap_1, \heap_2) \wDefined{\attacker}
\world$. These relations are all parametrized by the level of the
observer (adversary), $\attacker$, which is an element of $\lattice$.

The value relation $\Blrv {\tau}$ defines, for each type, which pairs
of values from the two runs are related by that type (at each world,
each step-index and each adversary). At base types, $\tbase$, only
identical values are related. For pairs, the relation is the intuitive
one: $(v_1,v_2)$ and $(v_1',v_2')$ are related in $\Blrv{\tau_1 \times
  \tau_2}$ iff $v_i$ and $v_i'$ are related in $\Blrv{\tau_i}$ for $i
\in \{1,2\}$. Two values are related at a sum type only if they are
both left injections or both right injections. At the function type
$\tau_1 \fto \tau_2$, two functions are related if they map values
related at the argument type $\tau_1$ to expressions related at the
result type $\tau_2$. For technical reasons, we also need both the
functions to satisfy the conditions of the \emph{unary} relation. At a
reference type $\tref~\tau$, two locations $a_1$ and $a_2$ are related
at world $\world = (\uWorld_1, \uWorld_2, \pb)$ only if they are
related by $\pb$ (i.e., they are correspondingly allocated locations)
and their types as specified by $\uWorld_1$ and $\uWorld_2$ are equal
to $\tau$.

Finally, and most importantly, at a labeled type $\utype^{\llabel}$,
$\Blrv{\utype^{\llabel}}$ relates values depending on the ordering
between $\llabel$ and the adversary $\attacker$. When $\llabel \lbelow
\attacker$, the adversary can see values labeled $\llabel$, so
$\Blrv{\utype^{\llabel}}$ contains exactly the values related in
$\Blrv{\utype}$. When $\llabel \not\lbelow \attacker$, values labeled
$\llabel$ are opaque to the adversary (in colloquial terms, they are
``high''), so they can be arbitrary. In this case,
$\Blrv{\utype^{\llabel}}$ is the cross product of the \emph{unary}
interpretation of $\utype$ with itself. This is the only place in our
model where the binary and unary interpretations interact.

The expression relation $\Blre {\tau}$ defines, for each type, which
pairs of expressions from the two executions lie in the type (at each
world $\world$, each step index $n$ and each
adversary~$\attacker$). The definition is similar to that in the unary
case: $e_1$ and $e_2$ lie in $\Blre{\tau}$ if the values they produce
are related in the value relation $\Blrv{\tau}$, and the expressions
preserve heap conformance.

The heap conformance relation $(n, \heap_1, \heap_2)
\wDefined{\attacker} \world$ defines when a pair of heaps $\heap_1$,
$\heap_2$ conforms to a world $\world=(\uWorld_1, \uWorld_2,
\pb)$. The relation requires that any pair of locations related by
$\pb$ have the same types (according to $\uWorld_1$ and $\uWorld_2$),
and that the values stored in $\heap_1$ and $\heap_2$ at these
locations lie in the binary value relation of that common type.

\medskip
\noindent \textbf{Meta-theory.}  The primary meta-theoretic property
of a logical relations model like ours is the so-called
\emph{fundamental theorem}. This theorem says that any expression
syntactically in a type (as established via the type system) also lies
in the semantic interpretation (the expression relation) of that
type. Here, we have two such theorems---one for the unary
interpretation and one for the binary interpretation.

To write these theorems, we define unary and binary interpretations of
contexts, $\Ulrv{\Gamma}$ and $\Blrv{\Gamma}$, respectively. These
interpretations specify when unary and binary substitutions conform to
$\Gamma$. A unary substitution $\vsubsts$ maps each variable to a
value whereas a binary substitution $\vsubstp$ maps each variable to
two values, one for each run.
\begin{mathpar}
  \begin{array}{@{}l@{\,}l@{\,}l@{}}
    \Ulrv {\Gamma} & \triangleq &
    \begin{array}[t]{@{}l@{}}
      \{(\uWorld, n, \vsubsts) ~|~  \dom(\Gamma) \subseteq \dom(\vsubsts) \wedge
                                  \forall x \in \dom(\Gamma). \\
~~~~~~~~~~~~~~~~(\uWorld, n, \vsubsts(x)) \in \Ulrv {\Gamma(x)}
       \}
    \end{array}\\
  \Blrv {\Gamma} & \triangleq &
  \begin{array}[t]{@{}l@{}}
    \{(\world, n, \vsubstp) ~|~
    \dom(\Gamma) \subseteq \dom(\vsubstp) \wedge \forall x \in
    \dom(\Gamma). \\
~~~~~~~~~~~~    (\world, n, \pi_1(\vsubstp(x)), \pi_2(\vsubstp(x)))
    \in \Blrv {\Gamma(x)} \}
    \end{array}
  \end{array}
\end{mathpar}

\begin{thm}[Unary fundamental theorem]
\label{thm:fundUnary-fg}
If $\Gamma \vdash_{\pc} e:\tau$
and $(\uWorld, n, \vsubsts) \in \Ulrv {\Gamma }$, then
$(\uWorld, n, e ~ \vsubsts) \in \lreu {\tau} {\pc}$.
\end{thm}

\begin{thm}[Binary fundamental theorem]
  \label{thm:fundBinary-fg}
  If $\Gamma \vdash_{\pc} e:\tau$ and $(\world, n, \vsubstp) \in \Blrv
  {\Gamma}$, then $(\world, n, e ~ (\vsubstp \proj{1}), e ~ (\vsubstp
  \proj {2})) \in \Blre {\tau} $, where $\vsubstp\proj{1}$ and
  $\vsubstp\proj{2}$ are the left and right projections of $\gamma$.
\end{thm}

The proofs of these theorems proceed by induction on the given
derivations of $\Gamma \vdash_{\pc} e:\tau$. The proofs are tedious,
but not difficult or surprising. The primary difficulty, as with all
logical relations models, is in setting up the model correctly, not in
proving the fundamental theorems.

{\fg}'s noninterference theorem (Theorem~\ref{thm:ni-fg}) is a simple
corollary of these two theorems.

%%% Local Variables:
%%% mode: latex
%%% TeX-master: "main"
%%% End:

\subsection{The coarse-grained type system, {\cg}}
\label{sec:cg}

\begin{figure*}[!htbp]
  % \textbf{Term, type, constraint syntax:}\\[-6pt]

\newcolumntype{L}{>{$}l<{$}}
\newcolumntype{P}{>{$}X<{$}}

\begin{tabularx}{\textwidth}{lLLP}
  Expressions & e & ::= & x \mid \lambda x.e \mid e \; e \mid (e, e) \mid %
	\efst(e) \mid \esnd(e) \mid \einl(e) \mid \einr(e) \mid %
	\ecase(e, x.e, y.e) \mid \enew e \mid {!e} \mid e := e \mid %
	() \mid %
	\elabel{}(e) \mid \eunlabel(e) \mid %
	\etolabeled(e) \mid \eret(e) \mid \ebind(e, x.e) %
	\\

  %% Labels & \llabel & ::= & \bot \mid \top \mid \lelement \mid \llabel \ljoin \llabel \mid
  %%       \llabel \lmeet \llabel\\

  Types & \tau & ::= & \tbase \mid \tunit \mid \tau \to \tau \mid \tau \times \tau \mid \tau + \tau
  \mid \tref~\llabel~\tau \mid \tlabeled{\llabel}{\tau} \mid
	\tslio {\llabel_1}{\llabel_2}{\tau} \\
\end{tabularx} \\ \\

\textbf{Typing judgment:}
\framebox{$\Gamma \vdash e: \tau$}\\

(All rules of the simply typed lambda-calculus pertaining to the types
$\tbase, \tau \to \tau, \tau \times \tau, \tau + \tau$, and $\tunit$
are included.)

\begin{mathpar}
   \inferrule
   { \Gamma \vdash e_1: \tslio{\llabel_1}{\llabel_2}{\tau}
     \\ \Gamma, x: \tau \vdash e_2 : \tslio {\llabel_3}{\llabel_4}{\tau'}
     \\ \llabel \lbelow \llabel_1
     \\ \llabel \lbelow \llabel_3
     \\ \llabel_2 \lbelow \llabel_3
     \\ \llabel_2 \lbelow \llabel_4 
%     \\ \llabel_4 \lbelow \llabel_4
   }
   {
    \Gamma \vdash \ebind(e_1, x.e_2)
   		: \tslio {\llabel} {\llabel_4} {\tau'}
   }
   \rname{{\cg}-bind}

   \inferrule
   { \Gamma \vdash e: \tau }
   { \Gamma \vdash \eret(e): \tslio {\top} {\bot} {\tau} }
   \rname{{\cg}-ret}

   \inferrule
   { \Gamma \vdash e: \tau'
   	\\ \lattice \vdash \tau' \subtype \tau
   }
   { \Gamma \vdash e: \tau }
   \rname{{\cg}-sub}

   \inferrule
   { \Gamma \vdash e: \tau
   	% \\ \lattice \vdash \llabel_i \lbelow \llabel
   }
   {
    \Gamma \vdash \elabel{}(e)
    : {\tlabeled{\llabel}{\tau}}
   }
   \rname{{\cg}-label}

   \inferrule
   { \Gamma \vdash e: \tlabeled{\llabel}{\tau} }
   {
    \Gamma \vdash \eunlabel(e)
   	: \tslio{\top}{\llabel}{\tau}
   }
   \rname{{\cg}-unlabel}

   \inferrule
   { \Gamma \vdash e : \tlabeled{\llabel}{\tau}
   }
   {
    \Gamma \vdash \enew e
    		: \tslio{\llabel}{\bot}{(\tref ~ {\llabel} ~ {\tau})}
   }
   \rname{{\cg}-ref}

   \inferrule
   { \Gamma \vdash e : \tref ~ \llabel' ~ \tau }
   {
    \Gamma \vdash {!e}
    		: \tslio{\top}{\bot}{(\tlabeled \llabel' \tau)}
   }
   \rname{{\cg}-deref}

   \inferrule
   { \Gamma \vdash e_1 : \tref ~{\llabel}~ \tau
   	\\ \Gamma \vdash e_2 : \tlabeled {\llabel}{\tau}
   }
   {
     \Gamma \vdash e_1 := e_2
     : \tslio{\llabel}{\bot}{\tunit}
   }
   \rname{{\cg}-assign}

   \inferrule
   { \Gamma \vdash e: \tslio{\llabel}{\llabel'}{\tau} }
   {
    \Gamma \vdash \etolabeled(e)
   		: \tslio{\llabel}{\bot}{(\tlabeled{\llabel'}{\tau})}
   }
   \rname{{\cg}-toLabeled}

  \end{mathpar}

  \textbf{Subtyping judgment:}
  \framebox{$\lattice \vdash \tau \subtype \tau'$}

  \begin{mathpar}
   \inferrule
   { \lattice \vdash \tau \subtype \tau'
   	\\ \lattice \vdash \llabel \lbelow \llabel'
   }
   {
    \lattice \vdash \tlabeled \llabel \tau \subtype \tlabeled {\llabel'} \tau'
   }
  \rname{{\cg}sub-labeled}

   \inferrule
   { \lattice \vdash \tau \subtype \tau'
   	\\ \lattice \vdash \llabel_1' \lbelow \llabel_1
   	\\ \lattice \vdash \llabel_2 \lbelow \llabel_2'
   }
   {
    \lattice \vdash \tslio{\llabel_1}{\llabel_2}{\tau}
   		\subtype \tslio{\llabel_1'}{\llabel_2'} {\tau'}
   }
   \rname{{\cg}sub-monad}
   \end{mathpar}
  %% \inferrule
  %% { }
  %% {
  %%   \lattice \vdash \tau \subtype \tau
  %% }\rname{{\cg}sub-refl}

  %%  \inferrule
  %%  { \lattice \vdash \tau_1' \subtype \tau_1
  %%  	\\ \lattice \vdash \tau_2 \subtype \tau_2' \
  %%  }
  %%  { \lattice \vdash \tau_1 \to \tau_2 \subtype \tau_1' \to \tau_2' }
  %% \rname{{\cg}sub-arrow}

  %%  \inferrule
  %%  { \lattice \vdash \tau_1 \subtype \tau_1'
  %%  	\\ \lattice \vdash \tau_2 \subtype \tau_2'
  %%  }
  %%  { \lattice \vdash \tau_1 \times \tau_2 \subtype \tau_1' \times \tau_2' }
  %% \rname{{\cg}sub-prod}

  %%  \inferrule
  %%  { \lattice \vdash \tau_1 \subtype \tau_1'
  %%  	\\ \lattice \vdash \tau_2 \subtype \tau_2'
  %%  }
  %%  { \lattice \vdash \tau_1 + \tau_2 \subtype \tau_1' + \tau_2' }
  %% \rname{{\cg}sub-sum}
  \caption{{\cg}'s language syntax and type system (selected rules)}
  \label{fig:cg-ts}
\end{figure*}

%% \begin{figure*}[!htbp]
%% \caption{{\cg} subtyping}
%% \label{fig:cg-subtyping}
%% \end{figure*}

%%% Local Variables:
%%% mode: latex
%%% TeX-master: "main"
%%% End:

Our coarse-grained type system, {\cg}, is a variant of the static
fragment of the hybrid IFC type system
HLIO~\cite{icfp15-HLIO}.\footnote{\update{Differences between {\cg} and HLIO
  and their consequences are discussed in
  Section~\ref{sec:discussion}.}} Like {\fg}, {\cg} also operates on a
  higher-order, eager, call-by-value language with state, but it
  separates pure expressions from impure (stateful) ones at the level
  of types. This is done, as usual, by introducing a monad for state,
  and limiting all state-accessing operations (dereferencing,
  allocation, assignment) to the monad. However, in {\cg}, as in HLIO,
  the monad also doubles as the unit of labeling. Values and types are
  not necessarily labeled individually in {\cg}. Instead, there is a
  confidentiality label on an entire monadic computation. This makes
  {\cg} coarse-grained.

{\cg}'s syntax and type system are shown in
Figure~\ref{fig:cg-ts}. The types include all the usual types of the
simply typed $\lambda$-calculus and, unlike {\fg}, a label is not
forced on every type. There are two special types: $\tslio
{\llabel_1}{\llabel_2}{\tau}$ and $\tlabeled{\llabel}{\tau}$.

The type $\tslio {\llabel_1}{\llabel_2}{\tau}$ is the aforementioned
monadic type of computations that may access the heap (expressions of
other types cannot access the heap), eventually producing a value of
type $\tau$. The first label $\llabel_1$, called the \emph{pc-label},
is a lower bound on the write effects of the computation. The second
label $\llabel_2$, called the \emph{taint label}, is an upper bound on
the labels of all values that the computation has analyzed so far; it
is, for this reason, also an \emph{implicit} label on the output type
$\tau$ of the computation, and on any intermediate values within the
computation.

The type $\tlabeled{\llabel}{\tau}$ explicitly labels a value (of type
$\tau$) with label~$\llabel$. In {\fg}'s notation, this would be
analogous to $\tau^\llabel$. The difference is that this labeling can
be used \emph{selectively} in {\cg}; unlike {\fg}, not every type must
be labeled. Also, the reference type $\tref~\llabel~\tau$ carries an
explicit label $\llabel$ in {\cg}. Such a reference stores a value of
type $\tlabeled{\llabel}{\tau}$. \update{Labels on references are
  necessary to prevent implicit leaks via control dependencies---the
  type system relates the pc-label to the label of the written value
  at every assignment. Similar labels on references were unnecessary
  in {\fg} since every written value always carries a label anyhow.}

\medskip
\noindent\textbf{Typing rules.}  {\cg} uses the standard typing
judgment $\Gamma \vdash e: \tau$. There is no need for a $\pc$ on the
judgment since effects are confined to the monad. {\cg} uses the
typing rules of the simply typed $\lambda$-calculus for the type
constructs $\tbase$, $\tunit$, $\times$, $+$ and $\rightarrow$. We do
not re-iterate these standard rules, and focus here only on the new
constructs. The construct $\eret(e)$ is the monadic return that
immediately returns $e$, without any heap access. Consequently, it can
be given the type $\tslio{\top}{\bot}{\tau}$ (rule
\rname{{\cg}-ret}). The pc-label is $\top$ since the computation has
no write effect, while the taint label is $\bot$ since the computation
has not analyzed any value.

The monadic construct $\ebind(e_1, x.e_2)$ sequences the computation
$e_2$ after $e_1$, binding the return value of $e_1$ to $x$ in
$e_2$. The typing rule for this construct, \rname{{\cg}-bind}, is
important and interesting. The rule says that $\ebind(e_1, x.e_2)$ can
be given the type $\tslio {\llabel} {\llabel_4} {\tau'}$ if $(e_1:
\tslio{\llabel_1}{\llabel_2}{\tau})$, $(e_2:
\tslio{\llabel_3}{\llabel_4}{\tau'})$ and four conditions hold. The
conditions $\llabel \lbelow \llabel_1$ and $\llabel \lbelow \llabel_3$
check that the pc-label of $\ebind(e_1, x.e_2)$, which is $\llabel$,
is below the pc-label of $e_1$ and $e_2$, which are $\llabel_1$ and
$\llabel_3$, respectively. This ensures that the write effects of
$\ebind(e_1, x.e_2)$ are indeed above its pc-label, $\llabel$. The
conditions $\llabel_2 \lbelow \llabel_3$ and $\llabel_2 \lbelow
\llabel_4$ prevent leaking the output of $e_1$ via
the write effects and the output of $e_2$, respectively. Observe how
these conditions together track labels at the level of entire
computations, i.e., coarsely.

Next, we describe rules pertaining to the type
$\tlabeled{\llabel}{\tau}$. This type is introduced using the
expression constructor $\elabel{}$, as in rule
\rname{{\cg}-label}. Dually, if $e: \tlabeled{\llabel}{\tau}$, then
the construct $\eunlabel(e)$ eliminates this label. This construct has
the monadic type $\tslio{\top}{\llabel}{\tau}$. The taint label
$\llabel$ indicates that the computation has (internally) analyzed
something labeled $\llabel$. The pc-label is $\top$ since nothing has
been written.

Rule \rname{{\cg}-deref} says that dereferencing (reading) a location
of type $\tref~\llabel'~\tau$ produces a computation of type
$\tslio{\top}{\bot}{(\tlabeled{\llabel'}{\tau})}$. The type is monadic
because dereferencing accesses the heap. The value the computation
returns is explicitly labeled at $\llabel'$. The pc-label is $\top$
since the computation does not write, while the taint label is $\bot$
since the computation does not analyze the value it reads from the
reference. (The taint label will change to $\llabel'$ if the read value
is subsequently unlabeled.) Dually, the rule \rname{{\cg}-assign}
allows assigning a value labeled $\llabel$ to a reference labeled
$\llabel$. The result is a computation of type
$\tslio{\llabel}{\bot}{\tunit}$. The pc-label $\llabel$ indicates a
write effect at level $\llabel$.

The last typing rule we highlight pertains to a special construct,
$\etolabeled(e)$. This construct transforms $e$ of monadic type
$\tslio{\llabel}{\llabel'}{\tau}$ to the type
$\tslio{\llabel}{\bot}{(\tlabeled{\llabel'}{\tau})}$. This is
perfectly safe since the only way to observe the output of a monad is
by binding its result and that result is explicitly labeled in the
final type. The purpose of using this construct is to reduce the taint
label of a computation to $\bot$. This allows a subsequent
computation, which will \emph{not} analyze the output of the current
computation, to avoid raising its own taint label to
$\llabel'$. Hence, this construct limits the scope of the taint label
to a single computation, and prevents overtainting subsequent
computations. We make extensive use of this construct in our
translation from {\fg} to {\cg}. We note that HLIO's original typing
rule for $\etolabeled$ is different, and does not always allow
reducing the taint to $\bot$. We discuss the consequences of this
difference in Section~\ref{sec:discussion}.

We briefly comment on subtyping for specific constructs in
{\cg}. Subtyping of $\tlabeled{\llabel}{\tau}$ is co-variant in
$\llabel$, since it is always safe to increase a confidentiality
label. Subtyping of $\tslio{\llabel_1}{\llabel_2}{\tau}$ is
contra-variant in the pc-label $\llabel_1$ and co-variant in the taint
label $\llabel_2$ since the former is a lower bound while the latter
is an upper bound.

We prove soundness for {\cg} by showing that every well-typed
expression satisfies noninterference. Due to the presence of monadic
types, the soundness theorem takes a specific form (shown below), and
refers to a \emph{forcing semantics}. These semantics operate on
monadic types and actually perform reads and writes on the heap (in
contrast, the pure evaluation semantics simply return suspended
computations for monadic types). The forcing semantics are the
expected ones, so we defer their details to the appendix.

\begin{thm}[Noninterference for {\cg}]
  \label{thm:ni-cg}
  Suppose (1) $\llabel_i \not\lbelow \llabel$, (2)
  $x: \tlabeled{\llabel_i}{\tau} \vdash e: \tslio{\_}{\llabel}{\tbool}$, and (3)
  $v_1, v_2: \tlabeled{\llabel_i}{\tau}$. If both $e[v_1/x]$ and $e[v_2/x]$ terminate when forced,
  then they produce the same value (of type $\tbool$).
\end{thm}

\subsubsection{Semantic model of {\cg}}
We build a semantic model of {\cg}'s types. The model is very similar
in structure to the model of {\fg}'s types. We use two
interpretations, unary and binary, and worlds exactly as in {\fg}'s
model. The difference is that since state effects are confined to a
monad in {\cg}, all the constraints on heap updates move from the
expression relations to the value relations at the monadic
types. Owing to lack of space, and the similarity in the structures of
the models, we defer {\cg}'s model and the fundamental theorems to the
appendix.

%%% Local Variables:
%%% mode: latex
%%% TeX-master: "main"
%%% End:

%%% Local Variables:
%%% mode: latex
%%% TeX-master: "main"
%%% End:

%\input{semantic-model}
\section{Translations}
\label{sec:trans}

In this section, we describe our translations from {\fg} to {\cg} and
vice-versa, thus showing that these two type systems are equally
expressive. We start with the translation from {\fg} to {\cg}.

%% In this section we give a type-preserving translation from {\fg} to {\cg}, the missing direction
%% from~\cite{siglog17-ifcComp}. To explain this, we first recap the key elements of the translation from {\rfg}
%% (a fragment of {\fg}) to {\cg} from~\cite{siglog17-ifcComp} in Section~\ref{sec:recap}. Then we describe a
%% type-preserving translation from {\fg} to {\rfg} (in Section~\ref{sec:fg-2-rfg}) as a stepping stone
%% to achieve the full translation from {\fg} to {\cg}. And finally we use the insights from {\fg} to
%% {\rfg} translation to achieve a direct translation from the whole of {\fg} to {\cg} in
%% Section~\ref{sec:fg-2-cg}.

%% \input{recap}

%% \input{fg-to-rfg}

\subsection{Translating {\fg} to {\cg}}
\label{sec:fg-2-cg}

\begin{figure*}[!htbp]
\begin{mathpar}
  \inferrule
  { }
  {
    \Gamma, x:\tau \vdash_{\pc} x:\tau \leadsto \eret ~ x
  }\rname{\fc-var}

  \inferrule
  {
    \Gamma, x:\tau_1 \vdash_{\llabel_e} e: \tau_2 \leadsto e_{c1}
  }
  {
    \Gamma \vdash_{\pc} \lambda x. e: (\tau_1 \fto \tau_2)^{\bot}
    \leadsto
    \eret (\elabel {} (\lambda x. e_{c1}))
  }\rname{\fc-lam}

  \inferrule
  {
    \Gamma \vdash_{\pc} e_1: (\tau_1 \fto \tau_2)^{\llabel}
    \leadsto
    e_{c1}
    \\
    \Gamma \vdash_{\pc} e_2:\tau_1
    \leadsto
    e_{c2}
    \\
    \lattice \vdash  \llabel \ljoin \pc \lbelow \llabel_e \\
    \lattice \vdash \tau_2 \searrow \llabel
  }
  {
    \Gamma \vdash_{\pc} e_1 ~ e_2:\tau_2
    \leadsto
    \coerce(\ebind(e_{c1}, a. \ebind (e_{c2}, b. \ebind(\eunlabel ~ a, c. (c ~ b)))))
  }\rname{\fc-app}

  \inferrule
  {
    \Gamma \vdash_{\pc} e_1:\tau_1 \leadsto e_{c1}
    \\
    \Gamma \vdash_{\pc} e_2:\tau_2 \leadsto e_{c2}
  }
  {
    \Gamma \vdash_{\pc} (e_1, e_2): (\tau_1 \times \tau_2)^{\bot}
    \leadsto
    \ebind(e_{c1}, a. \ebind (e_{c2}, b. \eret(\elabel {} (a,b))))
  }\rname{\fc-prod}

  \inferrule
  {
    \Gamma \vdash_{\pc} e: (\tau_1 \times \tau_2)^{\llabel} \leadsto e_{c}
    \\
    \lattice \vdash \tau_1 \searrow \llabel
  }
  {
    \Gamma \vdash_{\pc} \efst (e):\tau_1
    \leadsto
    \coerce(\ebind(e_{c}, a. \ebind(\eunlabel ~ a, b. \eret (\efst (b)))))
  }\rname{\fc-fst}

  %% \inferrule
  %% {
  %%   \Gamma \vdash_{\pc} e: (\tau_1 \times \tau_2)^{\llabel} \leadsto e_{c}
  %%   \\
  %%   \lattice \vdash \tau_2 \searrow \llabel
  %% }
  %% {
  %%   \Gamma \vdash_{\pc} \esnd (e):\tau_2
  %%   \leadsto
  %%   \coerce(\ebind(e_{c}, a. \ebind(\eunlabel ~ (a), b. \eret (\esnd (b)))))
  %% }\rname{\fc-snd}

  \inferrule
  {
    \Gamma \vdash_{\pc} e:\tau_1 \leadsto e_{c}
  }
  {
    \Gamma \vdash_{\pc} \einl (e):(\tau_1 + \tau_2)^{\bot}
    \leadsto
    \ebind(e_{c}, a. \eret (\elabel {} {(\einl (a))}))
  }\rname{\fc-inl}

  %% \inferrule
  %% {
  %%   \Gamma \vdash_{\pc} e:\tau_2 \leadsto e_{c}
  %% }
  %% {
  %%   \Gamma \vdash_{\pc} \einr (e):(\tau_1 + \tau_2)^{\bot}
  %%   \leadsto
  %%   \ebind(e_{c}, a. \eret (\elabel {} {(\einr (a))}))
  %% }\rname{\fc-inr}

  \inferrule
  {
    \Gamma \vdash_{\pc} e:(\tau_1 + \tau_2)^{\llabel} \leadsto e_{c}
    \\
    \Gamma, x:\tau_1 \vdash_{\pc \ljoin \llabel} e_1:\tau \leadsto e_{c1}
    \\
    \Gamma, x:\tau_1 \vdash_{\pc \ljoin \llabel} e_2:\tau \leadsto e_{c2}
    \\
    \lattice \vdash \tau \searrow \llabel
  }
  {
    \Gamma \vdash_{\pc} \ecase (e, x.e_1, y.e_2):\tau
    \leadsto
    \coerce(\ebind(e_{c}, a. \ebind(\eunlabel ~ a,  b. \ecase(b, x.e_{c1}, y. e_{c2}))))
  }\rname{\fc-case}

  %% \inferrule
  %% {
  %%   \Gamma \vdash_{\pc} e: \tau \leadsto e_c
  %%   \\
  %%   \lattice \vdash \tau \searrow \pc
  %% }
  %% {
  %%   \Gamma \vdash_{\pc} \enew (e): (\tref ~ \tau)^\bot
  %%   \leadsto
  %%   \ebind(e_{c}, a. \ebind(\enew (a), b. \eret (\elabel {} {b})))
  %% }\rname{\fc-ref}

  \inferrule
  {
    \Gamma \vdash_{\pc} e: (\tref ~ \tau)^{\llabel} \leadsto e_c
    \\ \lattice \vdash \tau \subtype \tau'
    \\ \lattice \vdash \tau' \searrow \llabel
  }
  {
    \Gamma \vdash_{\pc} !e: \tau
    \leadsto
    \coerce(\ebind(e_{c}, a. \ebind(\eunlabel ~ a, b. !b)))
  }\rname{\fc-deref}

  \inferrule
  {
    \Gamma \vdash_{\pc} e_1: (\tref ~ \tau)^{\llabel} \leadsto e_{c1}
    \\
    \Gamma \vdash_{\pc} e_2: \tau \leadsto e_{c2}
    \\
    \tau \searrow (\pc \ljoin \llabel)
  }
  {
    \Gamma \vdash_{\pc} e_1 := e_2: \tunit
    \leadsto
    % \etolabeled(\ebind(e_{c1}, a. \ebind(e_{c2}, b. \ebind (\eunlabel ~ a, c. c := b))))
    \ebind(\etolabeled(\ebind(e_{c1}, a. \ebind(e_{c2}, b. \ebind (\eunlabel ~ a, c. c := b)))), d. \eret ())
  }\rname{\fc-assign}
\end{mathpar}

\bigskip
\centering
where,
\framebox{
  \begin{tabular}[t]{l@{\ }c@{\ }l}
    $\coerce$ & $:$ & $\tslio{\pc}{\llabel}{\tau} \to \tslio{\pc}{\bot}{\tau}$ ~~~ when $\tau = \tlabeled{\llabel'}{\tau'}$ and $\llabel \lbelow \llabel'$ \\
    $\coerce$ & $\triangleq$ & $\lambda x. \etolabeled(\ebind(x, y.\eunlabel ~ y))$
  \end{tabular}
  }
\caption{Expression translation {\fg} to {\cg} (selected rules only)}
\label{fig:fg-2-cg-expr}
\end{figure*}

%%%Local Variables:
%%% mode: latex
%%% TeX-master: "main"
%%% End:

Our goal in translating {\fg} to {\cg} is to show how a fine-grained
IFC type system can be simulated in a coarse-grained one. Our
translation is directed by the type derivations in {\fg} and preserves
typing and semantics. We describe the translation below, followed by
formal properties of the translation. As a convention, we use the
subscript or superscript $s$ to indicate source ({\fg}) elements, and
$t$ to indicate target ({\cg}) elements. Thus $e_s$ denotes a source
expression, whereas $e_t$ denotes a target expression.

The key idea of our translation is to map a source expression $e_s$
satisfying $\vdash_{\pc} e_s: \tau$ to a monadic target expression
$e_t$ satisfying $\vdash e_t: \tslio{\pc}{\bot}{\transFC{\tau}}$. The
$\pc$ used to type the source expression is mapped as-is to the
pc-label of the monadic computation. The type of the source
expression, $\tau$, is translated by the function $\transFC{\cdot}$
that is described below. However---and this is the crucial bit---the
taint label on the translated monadic computation is $\bot$. To get
this $\bot$ taint we use the $\etolabeled$ construct judiciously. Not
setting the taint to $\bot$ can cause a taint explosion in translated
expressions, which would make it impossible to simulate the
fine-grained dependence tracking of {\fg}.

The function $\transFC{\cdot}$ defines how the types of source values
are translated. This function is defined by induction on labeled and
unlabeled source types.

\[
  \begin{array}{lll}
      \transFC{\tbase} & = & \tbase \\

      \transFC{\tunit} & = & \tunit\\

      \transFC{\tau_1 \overset{\llabel_e}{\to} \tau_2} & = &
                                                             \transFC{\tau_1} \to \tslio{\llabel_e}{\bot}{\transFC{\tau_2}} \\

      \transFC{\tau_1 \times \tau_2} & = & \transFC{\tau_1} \times \transFC{\tau_2} \\

      \transFC{\tau_1 + \tau_2} & = & \transFC{\tau_1} + \transFC{\tau_2} \\

      \transFC{\tref ~\tau} & = & \tref ~\llabel ~\transFC{\utype}~~~~\mbox{ when $\tau = \utype^{\llabel}$}\\

      \transFC{\utype^{\llabel}} & = & \tlabeled{\llabel}{\transFC{\utype}}
  \end{array}
\]
The translation should be self-explanatory. The only nontrivial case
is the translation of the function type $\tau_1
\overset{\llabel_e}{\to} \tau_2$. A source function of this type is
mapped to a target function that takes an argument of type
$\transFC{\tau_1}$ and returns a monadic computation (the translation
of the body of the source function) that has pc-label $\llabel_e$ and
eventually returns a value of type $\transFC{\tau_2}$.

Given this translation of types, we next define a type
derivation-directed translation of expressions. This translation is
formalized by the judgment $\Gamma \vdash_{\pc} e_s:\tau \leadsto
e_t$. The judgment means that translating the source expression $e_s$,
which has the typing derivation $\Gamma \vdash_{\pc} e_s$, yields the
target expression $e_t$. This judgment is \emph{functional}: For each
type derivation $\Gamma \vdash_{\pc} e_s: \tau$, it yields exactly one
$e_t$.  It is also easily implemented by induction on typing
derivations. The rules for the judgment are shown in
Figure~\ref{fig:fg-2-cg-expr}. The thing to keep in mind while reading
the rules is that $e_t$ should have the type
$\tslio{\pc}{\bot}{\transFC{\tau}}$.

We illustrate how the translation works using one rule,
\rname{\fc-app}. In this rule, we know inductively that the
translation of $e_1$, i.e., $e_{c1}$ has type
$\tslio{\pc}{\bot}{\transFC{(\tau_1 \overset{\llabel_e}{\to}
    \tau_2)^\llabel}}$, which is equal to
$\tslio{\pc}{\bot}{(\tlabeled{\llabel}{(\transFC{\tau_1} \to
    \tslio{\llabel_e}{\bot}{\transFC{\tau_2}})})}$. The translation of
$e_2$, i.e., $e_{c2}$ has type
$\tslio{\pc}{\bot}{\transFC{\tau_1}}$. We wish to construct something
of type $\tslio{\pc}{\bot}{\transFC{\tau_2}}$.

To do this, we bind $e_{c1}$ to the variable $a$, which has the type
$\tlabeled{\llabel}{(\transFC{\tau_1} \to
  \tslio{\llabel_e}{\bot}{\transFC{\tau_2}})}$. Similarly, we bind
$e_{c2}$ to the variable $b$, which has the type
$\transFC{\tau_1}$. Next, we unlabel $a$ and bind the result to
variable $c$, which has the type $\transFC{\tau_1} \to
\tslio{\llabel_e}{\bot}{\transFC{\tau_2}}$. However, due to the
unlabeling, the \emph{taint label on whatever computation we sequence
  after this bind must be at least $\llabel$}. Next, we apply $b$ to
$c$, which yields a value of type
$\tslio{\llabel_e}{\bot}{\transFC{\tau_2}}$. Via subtyping, using the
assumption $\pc \lbelow \llabel_e$, we can weaken this to
$\tslio{\pc}{\llabel}{\transFC{\tau_2}}$. This satisfies the
constraint that the taint label be at least $\llabel$ and is
\emph{almost} what we need, except that we need the taint $\bot$ in
place of $\llabel$.

To reduce the taint back to $\bot$, we use the \emph{defined} {\cg}
function $\coerce$, which has the type $\tslio{\pc}{\llabel}{\tau} \to
\tslio{\pc}{\bot}{\tau}$, when $\tau$ has the form
$\tlabeled{\llabel'}{\tau'}$ with $\llabel \lbelow \llabel'$. This
last constraint is satisfied here since we know that $\tau_2 \searrow
\llabel$. The function $\coerce$ uses $\etolabeled$ internally and is
defined in the figure.

This pattern of using $\coerce$, which internally contains
$\etolabeled$, to restrict the taint to $\bot$ is used to translate
all elimination forms (application, projection, case, etc.). Overall,
our translation uses $\etolabeled$ judiciously to prevent taint from
exploding in the translated expressions.

\textit{Remark.} Readers familiar with monads may note that our
translation from {\fg} to {\cg} is based on the standard
interpretation of the call-by-value $\lambda$-calculus in the
computational $\lambda$-calculus~\cite{moggi91:notions}. Our
translation additionally accounts for the pc and security labels, but
is structurally the same.

\begin{figure*}[!htbp]
  \centering
  \begin{displaymath}
    \begin{array}{lll}
      \fcUlrvn  {\tbase} & \triangleq & \{(\uWorlds, m,  \sval, \tval ) ~|~ \sval \in \llbracket \tbase \rrbracket \wedge
                                        \tval \in \llbracket \tbase \rrbracket \wedge \sval = \tval
                                        \} \\

%      \fcUlrvn  {\tunit} & \triangleq & \{(\uWorlds, m,  \sval, \tval) ~|~ \sval \in \llbracket \tunit \rrbracket
%                                        \wedge \tval \in \llbracket \tunit \rrbracket  \} \\

%      \fcUlrvn  {\tau_1 \times \tau_2} & \triangleq & \{(\uWorlds, m, (\sval_1,\sval_2), (\tval_1,\tval_2)) ~|~ \\
%                       & & (\uWorlds, m, \sval_1, \tval_1) \in \fcUlrvn  {\tau_1} \wedge
%                           (\uWorlds,  m,  \sval_2, \tval_2) \in \fcUlrvn  {\tau_2}
%                           \}\\

%      \fcUlrvn  {\tau_1 + \tau_2} & \triangleq & \{(\uWorlds,  m,  \einl ~  \sval, \einl ~ \tval) ~|~
%                                                 (\uWorlds,  m,  \sval, \tval) \in \fcUlrvn  {\tau_1} \} ~ \union \\
%                       & & \{(\uWorlds,  m,  \einr ~  \sval, \einr ~ \tval) ~|~
%                           (\uWorlds,  m,  \sval, \tval) \in \fcUlrvn  {\tau_2} %\} \\

      \fcUlrvn  {\tau_1 \fto \tau_2} & \triangleq & \{(\uWorlds,  m,  \lambda x. e_s, \lambda x. e_t)
                                                    ~|~ 
                        \forall \uWorlds' \wExtendsR \uWorlds,  \sval, \tval, j<m, \pb \wExtends \pb'.
                           (\uWorlds', j, \sval, \tval) \in \fclrvun {\tau_1} {\pb'}  \implies \\
                       & & \hspace{55mm} (\uWorlds', j, e_s[\sval/x], e_t[\tval/x]) \in \fclreun  {\tau_2} {\pb'}  \} \\

      \fcUlrvn  {\tref ~ \tau} & \triangleq & \{(\uWorlds, m, \loc_s, \loc_t) ~|~
                                              \uWorlds (\loc_s) = \tau \wedge ({^s}\loc, {^t}\loc) \in \pb \} \\

      \fcUlrvn  {\utype^{\llabel'}} & \triangleq & \{(\uWorlds, m, \sval, \elabel {} (\tval)) ~|~
                                                   (\uWorlds, m, \sval, \tval) \in
                                                   \fclrvun {\utype} {\pb}    \} \\ \\
%%%%%%%%%%%%%%%%%%%%%%%%%%%%%%%%%%%%%%%%%%%%%%
\hline \\                                                   
    \fcUlren {\tau} & \triangleq & \{(\uWorlds, n,  e_s, e_t) ~|~ 
                     \forall \heap_s, \heap_t. (n, \heap_s, \heap_t)
                        \wDefined{\pb} \uWorlds \wedge
                        \forall i<n, \sval. (\heap_s, e_s) \ReducesTo{i} (\heap_s',\sval) \implies \\
                    & & \hspace{24mm}\exists \heap_t', \tval. (\heap_t, e_t) \ReducesToF{} (\heap_t',\tval) \wedge
                        \exists  \uWorlds' \wExtendsR \uWorlds, \pb' \wExtendsR \pb. (n-i,\heap_s', \heap_t')
                        \wDefined{\pb'} \uWorlds' \\
                    & & \hspace{24mm}\wedge (\uWorlds', n-i, \sval, \tval) \in \fclrvun {\tau} {\pb'}
                        \}                                          
  \end{array}
\end{displaymath}
\caption{Cross-language value and expression relations for the {\fg} to {\cg} translation (excerpt)}
\label{fig:crossLangfg2cg}
\end{figure*}

\medskip\noindent \textbf{Properties.}  Our translation preserves
typing by construction. This is formalized in the following
theorem. The context translation $\transFC{\Gamma}$ is defined
pointwise on all types in~$\Gamma$.
\begin{thm}[Typing preservation]\label{thm:fg-2-cg-typePres}
  If $\Gamma \vdash_\pc e_s: \tau$ in {\fg}, then there is a unique
  $e_t$ such that $\Gamma \vdash_\pc e_s: \tau \leadsto e_t$ and that
  $e_t$ satisfies $\transFC{\Gamma} \vdash e_t:
  \tslio{\pc}{\bot}{\transFC{\tau}}$ in {\cg}.
\end{thm}
An immediate corollary of this theorem is that well-typed source
programs translate to noninterfering target programs (since target
typing implies noninterference in the target).

Next, we show that our translation preserves the meaning of programs,
i.e., it is semantically ``correct''. For this, we define a
\emph{cross-language} logical relation, which relates source values
(expressions) to target values (expressions) at each source type. This
relation has three key properties: (A) A source expression and its
translation are always in the relation
(Theorem~\ref{thm:fg2cg-FundThm}), (B) Related expressions reduce to
related values, and (C) At base types, the relation is the
identity. Together, these imply that our translation preserves the
meanings of programs in the sense that a function from base types to
base types maps to a target function with the same extension.

An excerpt of the relation is shown in
Figure~\ref{fig:crossLangfg2cg}. The relation is defined over source
({\fg}) types, and is divided (like our earlier relations) into a
value relation$\fcUlrvn{\cdot}$, an expression relation
$\fcUlren{\cdot}$, and a heap relation $(n, \heap_s, \heap_t)
\wDefined{\pb} \uWorlds$, which we omit here. The relations specify
when a source value (resp.\ expression, heap) is related to a target
value (resp.\ expression, heap) at a source unary world $\uWorlds$, a
step index $n$ and a partial bijection $\pb$ that relates source
locations to corresponding target locations. The relation actually
mirrors the unary logical relation for {\fg}. The definition of the
expression relation forces property (B) above, while the value
relation at base types forces property (C).

Our main result is again a fundamental theorem, shown below. The
symbols $\vsubstsS$ and $\vsubstsT$ denote unary substitutions in the
source and target, respectively. The relation $\fcUlrvn {\Gamma }$ is
the obvious one, obtained by pointwise lifting of the value relation;
its definition is in the appendix.

%% We close of the open terms by defining appropriate notion of related
%% substitutions like before but this time for a cross-language setting
%% (denoted by $\vsubsts$ described in appendix) and use that prove the
%% Fundamental theorem.

\begin{thm}[Fundamental theorem]\label{thm:fg2cg-FundThm}
If
  $\Gamma \vdash_{\pc} e_s:\tau \leadsto e_t$ 
and
  $(\uWorlds, n, \vsubstsS, \vsubstsT) \in \fcUlrvn  {\Gamma }$,
then
  $(\uWorlds, n, e_s ~ \vsubstsS, e_t ~ \vsubstsT) \in \fclreun {\tau } {\pb}$.
\end{thm}

The proof of this theorem is by induction on the derivation of $\Gamma
\vdash_{\pc} e_s:\tau \leadsto e_t$. This theorem has two important
consequences. First, it immediately implies property (A) above and,
hence, completes the argument that our translation is semantically
correct. Second, the theorem, along with the binary fundamental
theorem for {\cg}, allows us to re-derive the noninterference theorem
for {\fg} (Theorem~\ref{thm:ni-fg}) directly. This re-derivation is
important because it provides confidence that our translation
preserves the meaning of security labels. As a simple counterexample,
it is perfectly possible to translate {\fg} programs to {\cg}
programs, preserving both typing and semantics, by mapping all source
labels to the same target label (say, $\bot$). However, we would not
be able to re-derive the source noninterference theorem using the
target's properties if this were the case.

%%% Local Variables:
%%% mode: latex
%%% TeX-master: "main"
%%% End:

%\input{soundness-translations}

\subsection{Translating {\cg} to {\fg}}
\label{sec:cg2fg}

\begin{figure*}[!htbp]
\begin{mathpar}
  \inferrule
    {
       \Gamma \vdash e : \tau \leadsto e_F
    }
    {
       \Gamma \vdash \elabel{\llabel}(e)
      : \tlabeled{\llabel}{\tau}
      \leadsto
      \einl(e_F)
    }\rname{label}

    \inferrule
    {
      \Gamma \vdash e : \tlabeled{\llabel}{\tau} \leadsto e_F
    }
    {
      \Gamma \vdash \eunlabel(e) : \tslio{\top}{\llabel}{\tau}
      \leadsto
      \lambda \_. e_F
    }\rname{unlabel}

    \inferrule
    {
      \Gamma \vdash e : \tslio{\llabel_1}{\llabel_2}{\tau} \leadsto e_F
    }
    {
      \Gamma \vdash \etolabeled(e)
      : \tslio{\llabel_1}{\bot}{(\tlabeled{\llabel_2}{\tau})}
      \leadsto
      \lambda \_. \einl(e_F ~ ())
    }\rname{toLabeled}

    \inferrule
    {
      \Gamma \vdash e : \tau \leadsto e_F
    }
    {
      \Gamma \vdash \eret(e) : \tslio{\top}{\bot}{\tau}
      \leadsto \lambda \_. \einl(e_F)
    }\rname{ret}

    \inferrule
    {
      \Gamma \vdash e_1 : \tslio{\llabel_1}{\llabel_2}{\tau} \leadsto e_{F1}
      \\ \Gamma, x : \tau \vdash e_2 : \tslio{\llabel_3}{\llabel_4}{\tau'} \leadsto e_{F2}
     \\ \llabel \lbelow \llabel_1
      \\ \llabel \lbelow \llabel_3
      \\ \llabel_2 \lbelow \llabel_3
      \\ \llabel_2 \lbelow \llabel_4
     % \\ \llabel_4 \lbelow \llabel_4
    }
    {
      \Gamma \vdash \ebind(e_1, x.e_2) : \tslio{\llabel}{\llabel_4}{\tau'}
      \leadsto \lambda \_. \ecase(e_{F1} (), x. e_{F2} (), y.\einr())
    }\rname{bind}
  \end{mathpar}
    \caption{Expression translation {\cg} to {\fg} (selected rules only)}
  \label{fig:cg-to-fg-term}
\end{figure*}

This section describes the translation in the other direction---from
{\cg} to {\fg}. The overall structure (but not the details!) of this
translation are similar to that of the earlier {\fg} to {\cg}
translation, so we skip some boilerplate material here. The
superscript or subscript $s$ (source) now marks elements of {\cg} and
$t$ (target) marks elements of {\fg}.

The key idea of the translation is to map a source ({\cg}) expression
$e_s$ satisfying $\vdash e_s: \tau$ to a target ({\fg}) expression
$e_t$ satisfying $\vdash_\top e_t: \trans{\tau}$. The type translation
$\trans{\tau}$ is defined below. The $\pc$ for the translated
expression is $\top$ because, in {\cg}, all effects are confined to a
monad, so at the top-level, there are no effects. In particular, there
are no write effects, so we can pick any $\pc$; we pick the most
informative $\pc$, $\top$.

The type translation, $\trans{\tau}$, is defined by induction on
$\tau$.
\[
    \begin{array}{l@{~}c@{~}l}
      \trans{\tbase} 						& = & \tbase^\bot \\
      \trans{\tau_1 \to \tau_2} 			& = &
                                                              (\trans{\tau_1} \overset{\top}{\to} \trans{\tau_2})^\bot \\
      \trans{\tau_1 \times \tau_2} 		& = & (\trans{\tau_1} \times \trans{\tau_2})^\bot \\
      \trans{\tau_1 + \tau_2} 			& = & (\trans{\tau_1} + \trans{\tau_2})^\bot \\
  \trans{\tref~ \llabel~ \tau} 		& = & (\tref~ (\trans{\tau} + \tunit)^\llabel)^\bot \\
  \trans{\tslio{\llabel_1}{\llabel_2}{\tau}}
                                  & = &
                                        (\tunit \overset{\llabel_1}{\to} (\trans{\tau} + \tunit)^{\llabel_2})^\bot\\
  \trans{\tlabeled{\llabel}{\tau}} & = & (\trans{\tau} + \tunit)^\llabel
    \end{array}
\]

The most interesting case of the translation is that for
$\tslio{\llabel_1}{\llabel_2}{\tau}$. Since a {\cg} value of this type
is a suspended computation, we map this type to a \emph{thunk}---a
suspended computation implemented as a function whose argument has
type $\tunit$. The pc-label on the function matches the pc-label
$\llabel_1$ of the source type. The taint label $\llabel_2$ is placed
on the output type $\trans{\tau}$ using a coding trick: $(\trans{\tau}
+ \tunit)^{\llabel_2}$. The expression translation of monadic
expressions only ever produces values labeled $\einl$, so the right
type of the sum, $\tunit$, is never reached during the execution of a
translated expression. The same coding trick is used to translate
labeled and ref types. We could also have used a different coding in
place of $(\trans{\tau} + \tunit)^{\llabel_2}$. For example,
$(\trans{\tau} \times \tunit)^{\llabel_2}$ works equally well.

The expression translation is directed by source typing derivations
and is defined by the judgment $\Gamma \vdash e_s: \tau \leadsto e_t$,
some of whose rules are shown in Figure~\ref{fig:cg-to-fg-term}. The
translation is fairly straightforward (given the type
translation). The only noteworthy aspect is the use of the injection
$\einl$ wherever an expression of the type form $(\trans{\tau} +
\tunit)^{\llabel}$ needs to be constructed.

\medskip \noindent \textbf{Properties.}  The translation preserves
typing by construction, as formalized in the following theorem. The
context translation $\trans{\Gamma}$ is defined pointwise on all types
in $\Gamma$.
\begin{thm}[Typing preservation]\label{thm:cg-2-fg-typePres}
  If $\Gamma \vdash e_s: \tau$ in {\cg}, then there is a unique $e_t$
  such that $\Gamma \vdash e_s: \tau \leadsto e_t$ and that $e_t$
  satisfies $\trans{\Gamma} \vdash_\top e_t: \trans{\tau}$ in {\fg}.
\end{thm}
Again, a corollary of this theorem is that well-typed source programs
translate to noninterfering target programs.

We further prove that the translation preserves the semantics of
programs. Our approach is the same as that for the {\fg} to {\cg}
translation---we set up a cross-language logical relation, this time
indexed by {\cg} types, and show the fundamental theorem. From this,
we derive that the translation preserves the meanings of
programs. Additionally, we derive the noninterference theorem for
{\cg} using the binary fundamental theorem of {\fg}, thus gaining
confidence that our translation maps security labels properly. Since
this development mirrors that for our earlier translation, we defer
the details to the appendix.

\section{Discussion}
\label{sec:discussion}

\noindent \update{\textbf{Practical implications.}} \update{Our
  results establish that a coarse-grained IFC type system that labels
  at the granularity of entire computations can be as expressive as a
  fine-grained IFC type system that labels every individual value, if
  the coarse-grained type system has a construct like $\etolabeled$ to
  limit the scope of taints. It is also usually the case that a
  coarse-grained type system burdens a programmer less with
  annotations as compared to a fine-grained type system. This leads to
  the conclusion that, in general, there is merit to preferring
  coarse-grained IFC type systems with taint-scope limiting constructs
  over fine-grained IFC type systems. In a coarse-grained type system,
  the programmer can benefit from the reduced annotation burden and
  simulate the fine-grained type system when the fine-grained labeling
  is absolutely necessary for verification. Since our embedding of the
  fine-grained type system in the coarse-grained type system is
  compositional, it can be easily implemented in the coarse-grained
  type system as a library of macros, one for each construct of the
  language of the fine-grained type system.}

%% One can use macros
%% to simplify the top level expression translation. For instance, the
%% {\fc-app} rule from Figure~\ref{fig:fg-2-cg-expr}, can be simplified
%% as follows:

%% \begin{displaymath}
%%   \inferrule
%%   {
%%     \Gamma \vdash_{\pc} e_1: (\tau_1 \fto \tau_2)^{\llabel}
%%     \leadsto
%%     e_{c1}
%%     \\
%%     \Gamma \vdash_{\pc} e_2:\tau_1
%%     \leadsto
%%     e_{c2}
%%     \\
%%     \lattice \vdash  \llabel \ljoin \pc \lbelow \llabel_e \\
%%     \lattice \vdash \tau_2 \searrow \llabel
%%   }
%%   {
%%     \Gamma \vdash_{\pc} e_1 ~ e_2:\tau_2
%%     \leadsto
%%     \appCG \; e_{c1} \; e_{c2}
%%   }
%% \end{displaymath}

%% The simplified rule uses a macro $\appCG \; e_1' \; e_2'$ which is defined as

%% \noindent{$\coerce(\ebind(e_{1}', a. \ebind (e_{2}', b. \ebind(\eunlabel ~ a, c. (c ~ b)))))$}

\medskip
\noindent\textbf{Original HLIO.} The original HLIO
system~\cite{icfp15-HLIO}, from whose static fragment {\cg} is
adapted, differs from {\cg} in the interpretation of the labels
$\llabel_1$ and $\llabel_2$ in the monadic type
$\tslio{\llabel_1}{\llabel_2}{\tau}$. {\cg} interprets these labels as
the pc-label and the taint label, respectively. HLIO interprets these
labels as the \emph{starting taint} and the \emph{ending taint} of the
computation. This implies an \emph{invariant} that $\llabel_1 \lbelow
\llabel_2$ and makes HLIO more restrictive that {\cg}. The relevant
consequence of this difference is that the rule for $\etolabeled$
cannot always lower the final taint to $\bot$. HLIO's rule is:
\[   \inferrule
   { \Gamma \vdash e: \tslio{\llabel}{\llabel'}{\tau} }
   {
    \Gamma \vdash \etolabeled(e)
   		: \tslio{\llabel}{\llabel}{(\tlabeled{\llabel'}{\tau})}
   }
   \rname{{\cg}-toLabeled}
\]
This restrictive rule makes it impossible to translate from {\fg} to
HLIO in the way we translate from {\fg} to {\cg}. In
fact,~\cite{siglog17-ifcComp} already explains how this restriction
makes a translation from {\fg} to the static fragment of HLIO very
difficult. Our observation here is that HLIO's restriction, inherited
from a prior system called LIO, is not important for statically
enforced IFC and eliminating it allows a simple embedding of a
fine-grained IFC type system.

Nonetheless, we did investigate further whether we can embed {\fg}
into the static fragment of the unmodified HLIO. The answer is still
affirmative, but the embedding is complex and requires nontrivial
quantification over labels. Part II of our appendix contains a
complete account of this embedding (in fact, the appendix contains a
parallel account of all our results using the original HLIO in place
of {\cg}).

\update{HLIO also has two constructs, getLabel and labelOf, that allow
  reflection on labels. However, these constructs are meaningful only
  because HLIO uses hybrid (both static and dynamic) enforcement and
  carries labels at runtime. In a purely static enforcement, such as
  {\cg}'s, labels are not carried at runtime, so reflection on them is
  not meaningful.}

\medskip
\noindent \textbf{Full abstraction.}  Since our translations preserve
typed-ness, they map well-typed source programs to noninterfering
target programs. However, an open question is whether they preserve
contextual equivalence, i.e., whether they are fully
abstract. Establishing full abstraction will allow translated source
expressions to be freely co-linked with target expressions. We haven't
attempted a proof of full abstraction yet, but it looks like an
interesting next step. \update{We note that since our dynamic
  semantics (big-step evaluation) are not cognizant of IFC (which is
  enforced completely statically), it may be sufficient to generalize
  our translations to simply-typed variants of FG and CG, and prove
  those fully abstract.}

\medskip
\noindent \textbf{Other IFC properties.}  Our current setup is geared
towards proving \emph{termination-insensitive} noninterference, where
the adversary cannot observe nontermination. We believe that the
approach itself and the equivalence result should generalize to
termination-sensitive noninterference, but will require nontrivial
changes to our development. For example, we will have to change our
binary logical relations to imply co-termination of related
expressions and, additionally, modify the type systems to track
nontermination as a separate effect.

Another relevant question in whether our equivalence result can be
extended to type systems that support declassification and, more
foundationally, whether our logical relations can handle
declassification. This is a nuanced question, since it is unclear
hitherto how declassification can be given a compositional semantic
model. We are working on this problem currently.

%%% Local Variables:
%%% mode: latex
%%% TeX-master: "main"
%%% End:

\section{Related work}
\label{sec:related}

We focus on related work directly connected to our
contributions---logical relations for IFC type systems and language
translations that care about IFC.

\medskip \noindent \textbf{Logical relations for IFC type systems.}
Logical relations for IFC type systems have been studied before to a
limited extent. Sabelfeld and Sands develop a general theory of models
of information flow types based on partial-equivalence relations
(PERs), the mathematical foundation of logical
relations~\cite{esop99-PER-IFC}. However, they do not use these models
for proving any specific type system or translation sound. The pure
fragment of the SLam calculus was proven sound (in the sense of
noninterference) using a logical relations argument~\cite[Appendix
  A]{popl98-SLAM}. However, to the best of our knowledge, the relation
and the proof were not extended to mutable state. The proof of
noninterference for Flow Caml~\cite{toplas03-flowcaml}, which is very
close to SLam, considers higher-order state (and exceptions), but the
proof is syntactic, not based on logical relations. The dependency
core calculus (DCC)~\cite{popl99-DCC} also has a logical relations
model but, again, the calculus is pure. The DCC paper also includes a
state-passing embedding from the IFC type system of Volpano, Irvine
and Smith~\cite{jcs96-volpanoSmith}, but the state is
first-order. \update{Mantel \emph{et al.} use a security criterion
  based on an indistinguishability relation that is a PER to prove the
  soundness of a flow-sensitive type system for a concurrent
  language~\cite{DBLP:conf/csfw/MantelSS11}.  Their proof is also
  semantic, but the language is first-order.} In contrast to these
prior pieces of work, our logical relations handle higher-order state,
and this complicates the models substantially; we believe we are the
first to do so in the context of IFC.

Our models are based on the now-standard step-indexed Kripke logical
relations~\cite{DBLP:conf/popl/AhmedDR09}, which have been used
extensively for showing the soundness of program verification
logics. Our model for {\fg} is directly inspired by Cicek \emph{et
  al.}'s model for a pure calculus of incremental
programs~\cite{DBLP:conf/icfp/CicekP016}. That calculus does not
include state, but the model is structurally very similar to our model
of {\fg} in that it also uses a unary and a binary relation that
interact at labeled types. Extending that model with state was a
significant amount of work, since we had to introduce Kripke
worlds. Our model for {\cg} has no direct predecessor; we developed it
using ideas from our model of {\fg}. (DCC is also coarse-grained and
uses a labeled monad to track dependencies, but its model is quite
different from ours in the treatment of the monadic type.)

\medskip \noindent \textbf{Language translations that care about IFC.}
Language translations that preserve information flow properties appear
in the DCC paper. The translations start from SLam's pure fragment and
the type system of Volpano, Irvine and Smith and go into DCC. The
paper also shows how to recover the noninterference theorem of the
source of a translation from properties of the target, a theorem we
also prove for our translations.
Barthe \emph{et al.}~\cite{DBLP:journals/cl/BartheRB07} describe a
compilation from a high-level imperative language to a low-level
assembly-like language. They show that their compilation is type and
semantics preserving. They also derive noninterference for the source
from the noninterference of the target.
Fournet and
Rezk~\cite{popl08-cryptoIFC} describe a compilation from an IFC-typed
language to a low-level language where confidentiality and integrity
are enforced using cryptography. They prove that well-typed source
programs compile to noninterfering target programs, where the target
noninterference is defined in a computational sense.
Algehed and Russo~\cite{plas17-DCCinHaskell} define an embedding of
DCC into Haskell. They also consider an extension of DCC with state
but, to the best of our knowledge, they do not prove any formal
properties of the translation.

\section{Conclusion}
\label{sec:conclusion}

\update{This paper has examined the question of whether information
  flow type systems that label at fine granularity and those that
  label at coarse granularity are equally expressive.} We answer this
question in the affirmative, assuming that the coarse-grained type
system has a construct to limit the scope of the taint label. A more
foundational contribution of our work is a better understanding of
semantic models of information flow types. To this end, we have
presented logical relations models of types in both the fine-grained
and the coarse-grained settings, for calculi with mutable higher-order
state.

\medskip \noindent \textbf{\update{Acknowledgments.}}  \update{This
  work was partly supported by the German Science Foundation (DFG)
  through the project ``Information Flow Control for Browser Clients
  -- IFC4BC'' in the priority program ``Reliably Secure Software
  Systems -- RS$^3$'', and also through the Collaborative Research
  Center ``Methods and Tools for Understanding and Controlling
  Privacy'' (SFB 1223). We thank our anonymous reviewers and anonymous
  shepherd for their helpful feedback.}

%% In this paper we unified the two granularities of information flow tracking namely fine-grained and
%% coarse-grained. We extended a prior work \cite{siglog17-ifcComp} on this problem by coming up with a
%% type-preserving translation from {\fg} (a variant of FlowCaml) to {\cg} (a variant of HLIO), the
%% missing translation from \cite{siglog17-ifcComp}. We showed semantic soundness of the two type
%% systems involved and also proved that the type-preserving translations between them (in both
%% directions) preserve the program semantics and security of the source program. We are planning to
%% scale up this translation to other forms of dependence analysis like provenance tracking.

%%% Local Variables:
%%% mode: latex
%%% TeX-master: "main"
%%% End:

\bibliographystyle{IEEEtran}
\bibliography{citations}

\end{document}